\documentclass[letterpaper]{emulateapj}
\shorttitle{Small Planets Around Small Stars}
\shortauthors{Dressing \& Charbonneau}
\usepackage{epsfig}
\usepackage{amsmath}
\usepackage{rotating}
\usepackage{natbib}
\usepackage{enumerate}
\usepackage{hyperref}
\usepackage{url}
\usepackage{longtable}

\def\mearth{{\rm\,M_\oplus}}                                                    
\def\msun{{\rm\,M_\odot}}                                                       
\def\rsun{{\rm\,R_\odot}}                                                       
\def\rearth{{\rm\,R_\oplus}} 
\def\lsun{{\rm\,L_\odot}}                                                          
\def\kepler {{\emph{Kepler}\,}}                                              
\newcommand{\teff}{\ensuremath{T_{\mathrm{eff}}}}

\def\ndwarfs{3897 }
\def\ngiants{4420 }
\def\nambig{608 }

\def\nkoi{64}
\def\nhost{64}

\def\nplanets{95}
\def\nfitplanets{78} 

\def\npshrink{29\%}

\def\tsplit{3723}

\newcommand{\efifty}{\ensuremath{0.51_{-0.05}^{+0.06}} } 
\newcommand{\sefifty}{\ensuremath{0.39_{-0.04}^{+0.05}} }  
\newcommand{\totalseefifty}{\ensuremath{0.90^{+0.04}_{-0.03}} }  

\def\nhzsmall{2}

\newcommand{\ehz}{\ensuremath{0.15^{+0.13}_{-0.06}} } 

\newcommand{\sehz}{\ensuremath{0.04^{+0.06}_{-0.02}} } 

\def\hzocclim{0.04}

\def\hzoccbig{0.008}

\newcommand{\photmiss}{\ensuremath{21\%}} 

\def\nearlym{248 }
\def\nnewplanets{9}

\def\distlim{21}
\def\distlimnotransit{5}
 \def\distlimmostprob{13}
\def\distlimmostprobnotransit{3}
\begin{document}
\title{The Occurrence Rate of Small Planets around Small Stars}
\author{Courtney D. Dressing\altaffilmark{1,2}}
\author{David Charbonneau\altaffilmark{1}}
\altaffiltext{1}{Harvard-Smithsonian Center for Astrophysics, Cambridge, MA 02138}
\altaffiltext{2}{\tt cdressing@cfa.harvard.edu}
\vspace{0.5\baselineskip}
\date{\today}
\slugcomment{Accepted to ApJ}

\begin{abstract}
We use the optical and near-infrared photometry from the \kepler Input
Catalog to provide improved estimates of the stellar characteristics
of the smallest stars in the \kepler target list. We find \ndwarfs
dwarfs with temperatures below $4000$K, including \nkoi~planet
candidate host stars orbited by \nplanets~transiting planet candidates. 
We refit the transit events in the \kepler light curves for these planet 
candidates and combine the revised planet/star radius ratios with our
improved stellar radii to revise the radii of the planet candidates orbiting the
cool target stars. We then compare the number of observed planet
candidates to the number of stars around which such planets could have
been detected in order to estimate the planet occurrence rate around
cool stars. We find that the occurrence rate of $0.5-4\rearth$ planets
with orbital periods shorter than 50~days is \totalseefifty~planets per star. 
The occurrence rate of Earth-size ($0.5-1.4\rearth$) planets is constant 
across the temperature range of our sample at \efifty Earth-size planets 
per star, but the occurrence of 
$1.4-4\rearth$ planets decreases significantly at cooler temperatures. 
Our sample includes \nhzsmall~Earth-size planet
candidates in the habitable zone, allowing us to estimate that the
mean number of Earth-size planets in the habitable zone is
 \ehz planets per cool star. Our 95\% confidence
lower limit on the occurrence rate of Earth-size planets in the
habitable zones of cool stars is \hzocclim~planets per star. With 95\%
confidence, the nearest transiting Earth-size planet in the habitable
zone of a cool star is within \distlim~pc. Moreover, the nearest non-transiting planet in the habitable zone is within \distlimnotransit~pc with 95\% confidence.
\end{abstract}


\maketitle

\section{Introduction}
\label{sec:intro}
The \kepler mission has revolutionized exoplanet statistics by
increasing the number of known extrasolar planets and planet
candidates by a factor of five and discovering systems with longer orbital
periods and smaller planet radii than prior exoplanet surveys \citep{batalha_et_al2011,
borucki_et_al2012, fressin_et_al2012, gautier_et_al2012}. 
\kepler is a Discovery-class space-based mission designed 
to detect transiting exoplanets by monitoring the brightness 
of over 100,000~stars \citep{tenenbaum_et_al2012}.
The majority of \kepler's target stars are solar-like $FGK$ dwarfs and
accordingly most of the work on the planet occurrence rate from
\kepler has been focused on planets orbiting that sample of stars
\citep[e.g.,][]{borucki_et_al2011, cantanzarite+shao2011, youdin2011,
howard_et_al2012, traub2012}. Those studies revealed that the
planet occurrence rate increases toward smaller planet radii and
longer orbital periods. \citet{howard_et_al2012} also found evidence
for an increasing planet occurrence rate with decreasing stellar
effective temperature, but the trend was not significant below 5100K.

\citet{howard_et_al2012} conducted their analysis using the
1235~planet candidates presented in \citet{borucki_et_al2011}.  
The subsequent list of candidates published in February~2012 \citep{batalha_et_al2012} 
includes an additional 1091~planet
candidates and provides a better sample for estimating the occurrence
rate. The new candidates are primarily small objects (196 with $R_p <
1.25\rearth$, 416 with $1.25\rearth < R_p < 2\rearth$, and 421 with
$2\rearth < R_p < 6\rearth$), but the list also includes 41 larger
candidates with radii $6\rearth < R_p < 15\rearth$. The inclusion of
larger candidates in the \citet{batalha_et_al2012} sample is an
indication that the original \citet{borucki_et_al2011} list was not
complete at large planet radii and that continued improvements to
the detection algorithm may result in further announcements of planet
candidates with a range of radii and orbital periods.

In addition to nearly doubling the number of planet candidates,
\citet{batalha_et_al2012} also improved the stellar parameters for
many target stars by comparing the estimated temperatures, radii, and
surface gravities in the \kepler Input Catalog 
(KIC; \citealt{batalha_et_al2010b, brown_et_al2011}) to the values
expected from Yonsei-Yale evolutionary models
\citep{demarque_et_al2004}. Rather than refer back to the original
photometry, \citet{batalha_et_al2012} adopted the stellar parameters
of the closest Yonsei-Yale model to the original KIC values in the 
three-dimensional space of temperature, radius, and surface gravity.  
This approach did not correctly characterize the coolest target stars 
because the starting points were too far removed from the actual 
temperatures, radii, and surface gravities of the stars. In addition, 
the Yonsei-Yale models overestimate the observed radii and luminosity 
of cool stars at a given effective temperature \citep{boyajian_et_al2012}.

\subsection{The Small Star Advantage}
Although early work \citep{dole1964, kasting_et_al1993} 
suggested that a hypothetical planet in the habitable zone 
(the range of distances at which liquid water could exist on the 
surface of the planet) of an
M~dwarf would be inhospitable because the planet would be
tidally-locked and the atmosphere would freeze out on the dark side of
the planet, more recent studies have been more optimistic. For
instance, \citet{haberle_et_al1996} and \citet{joshi_et_al1997}
demonstrated that sufficient quantities of carbon dioxide could
prevent the atmosphere from freezing. In addition, \citet{pierrehumbert2011} 
reported that a tidally-locked planet could be in a partially habitable
``Eyeball Earth'' state in which the planet is mostly frozen but has a
 liquid water ocean at the substellar point.  Moreover, planets orbiting
M~dwarfs might become trapped in spin-orbit resonances like Mercury
instead of becoming spin-synchronized.

A second concern for the habitability of planets orbiting M~dwarfs is
the possibility of strong flares and high UV emission in quiescence \citep{france_et_al2012}.  Although a planet without a magnetic
field could require years to rebuild its ozone layer after
experiencing strong flare, the majority of the UV flux would never reach
the surface of the planet. Accordingly, flares do not present a
significant obstacle to the habitability of planets orbiting M dwarfs
\citep{segura_et_al2010}. Furthermore, the specific role of UV
radiation in the evolution of life on Earth is uncertain. A baseline
level of UV flux might be necessary to spur biogenesis
\citep{buccino_et_al2006}, yet UV radiation is also capable of
destroying biomolecules.

Having established that planets in the habitable zones of M~dwarfs
could be habitable despite the initial concern of the potential hazards 
of tidal-locking and
stellar flares, the motivation for studying the coolest target stars
is three-fold. First, several more years of \kepler observations will be
required to detect Earth-size planets in the habitable zones of
G~dwarfs due to the higher-than-expected photometric noise due to stellar variability
\citep{gilliland_et_al2011}, but \kepler is already sensitive to the
presence of Earth-size planets in the habitable zones of
M~dwarfs. Although a transiting planet in the habitable zone of a G
star transits only once per year, a transiting planet in the habitable
zone of a 3800K M~star transits five times per year. Additionally, the geometric
probability that a planet in the habitable zone transits the star is
1.8 times greater. Furthermore, the transit signal of an Earth-size
planet orbiting a 3800K M~star is 3.3~times deeper than the transit of an
Earth-size planet across a G star because the star is 45\% smaller
than the Sun. The combination of a shorter orbital period, an
increased transit probability, and a deeper transit depth greatly
reduces the difficulty of detecting a habitable planet and has
motivated numerous planet surveys to target M~dwarfs
\citep{delfosse_et_al1999, endl_et_al2003, nutzman+charbonneau2008,
zechmeister_et_al2009, apps_et_al2010, barnes_et_al2012, berta_et_al2012,
bowler_et_al2012, giacobbe_et_al2012, law_et_al2012}.

Second, as predicted by  \citet{salpeter1955} and \citet{chabrier2003}, 
studies of the solar neighborhood 
have revealed that M~dwarfs
are twelve times more abundant than
G~dwarfs. The abundance of M~dwarfs, 
combined with growing evidence for an increase
in the planet occurrence rate at decreasing stellar temperatures
\citep{howard_et_al2012}, implies that the majority of small planets
may be located around the coolest stars. Although M~dwarfs are
intrinsically fainter than solar-type stars, 75\% percent of the stars
within 10~pc are M~dwarfs\footnote{\url{http://www.recons.org/census.posted.htm}} \citep{henry_et_al2006}. 
These stars would be among the best targets for
future spectroscopic investigations of potentially-habitable rocky
planets due to the small radii and apparent brightness of the stars.

Third, confirming the planetary nature and measuring the mass of an
Earth-size planet orbiting within the habitable zone of an
M~dwarf is easier than confirming and measuring the mass of an 
Earth-size planet orbiting within the
habitable zone of a G~dwarf. The radial velocity signal induced by a
$1\mearth$ planet in the middle of the habitable zone ($a = 0.28$AU)
of a 3800K, $0.55\msun$ M~dwarf is $23$~cm/s. In comparison, the RV
signal caused by a $1\mearth$ planet in the habitable zone of a
G~dwarf is 9~cm/s. The prospects for RV~confirmation are even better
for planets around mid-to-late M~dwarfs: an Earth-size planet in the
habitable zone of a 3200K M~dwarf would produce an RV signal of
$1$~m/s, which is achievable with the current precision of modern
spectrographs \citep{dumusque_et_al2012}. Prior to investing a
significant amount of resources in investigations of the atmosphere of
a potentially habitable planet, it would be wise to first guarantee
that the candidate object is indeed a high-density planet and not a
low-density mini-Neptune.

Finally, upcoming facilities such as JWST and GMT will have the capability 
to take spectra of Earth-size planets in the habitable zones of M~dwarfs, 
but not Earth-size planets in the habitable zones of more massive stars.
In order to find a sample of habitable zone Earth-size planets for which astronomers 
could measure atmospheric properties with the next generation of 
telescopes, astronomers need to look for planets around small dwarfs.

\subsection{Previous Analyses of the Cool Target Stars}
In light of the advantages of searching for habitable planets around
small stars, several authors have worked on refining the parameters of
the smallest \kepler target stars. \citet{muirhead_et_al2012b}
collected medium-resolution, $K$-band spectra of the cool planet candidate host stars listed in \citet{borucki_et_al2011} and
presented revised stellar parameters for those host stars. Their sample
included 69~host stars with KIC~temperatures below 4400K as well as an
additional 13~host stars with higher KIC temperatures but with red colors that hint
that their KIC~temperatures were
overestimated. \citet{muirhead_et_al2012b} determined effective
temperature and metallicity directly from their spectra using the
H$_2$O-K2 index \citep{rojas-ayala_et_al2012} and then constrain
stellar radii and masses using Dartmouth stellar evolutionary models
\citep{dotter_et_al2008, feiden_et_al2011}. We adopt the same set of
stellar models in this paper. \citet{muirhead_et_al2012b} found that
one of the 82 targets (Kepler Object of Interest (KOI)~977) is a giant star and that three small
KOIs~(463.01, 812.03, 854.01) lie within the habitable zone.

\citet{johnson_et_al2012} announced the discovery of KOI~254.01, 
the first short-period gas giant orbiting an M~dwarf. 
The planet has a radius of $0.96 R_{\rm Jup}$ and 
orbits its host star KIC~5794240 once every 2.455239~days. 
In addition to discussing KOI~254.01, \citet{johnson_et_al2012} also
calibrated a relation for determining the masses and metallicities of
M~dwarfs from broad-band photometry. They found that $J-K$ color is a
reasonable ($\pm 0.15$~dex) indicator of metallicity for stars with
metallicities between $-0.5$ and 0.5~dex and $J-K$ colors within
0.1~magnitudes of the main sequence $J-K$ at the $V-K$ color of the
star in question. The relationship between infrared colors and metallicities was first proposed by  \citet{mould+hyland1976} and subsequently confirmed by \citet{leggett1992} and \citet{lepine_et_al2007}.

\citet{mann_et_al2012} took the first steps toward a global reanalysis
of the cool \kepler target stars. They acquired medium-resolution,
visible spectra of 382~target stars and classified all of the cool
stars in the target list as dwarfs or giants using ``training sets''
constructed from their spectra and literature
spectra. \citet{mann_et_al2012} found that the majority of bright, cool
target stars are giants in disguise and that the temperatures of the
cool dwarf stars are systematically overestimated by 110~K in the
KIC. \citet{mann_et_al2012} reported that
correctly classifying and removing giant stars removes the correlation
between cool star metallicity and planet occurrence observed by
\citet{schlaufman+laughlin2011}. After removing giant stars from the target list,
\citet{mann_et_al2012} calculated a planet occurrence rate of $0.37 \pm
0.08$ planets per cool star with radii between 2 and 32~$\rearth$ and
orbital periods less than 50~days. Their result is higher than the occurrence 
rate we report in Section~\ref{ssec:occrp}, most likely because of our 
revisions to the stellar radii.

In this paper, we characterize the coolest \kepler target stars by
revisiting the approach used to create the \kepler Input Catalog
\citep{brown_et_al2011} and tailoring that method for application to
cool stars. Specifically, we extract $grizJHK$ photometry from the KIC
for the 51813 planet search target stars with KIC~temperature
estimates $\leq 5050$K and for the 13402~planet search target stars
without KIC temperature estimates and compare the observed colors to
the colors of model stars from the Dartmouth Stellar Evolutionary
Database \citep{dotter_et_al2008, feiden_et_al2011}. We discuss the
features of the Dartmouth stellar models in
Section~\ref{ssec:dartmouth} and explain our procedure for assigning
revised stellar parameters in Section~\ref{ssec:assignments}. We
present revised stellar characterizations in Section \ref{sec:stars}
and improved planetary parameters for the associated planet candidates in Section
\ref{sec:planets}. We address the implications of these results on the
planet occurrence rate in Section \ref{sec:occ} and conclude in
Section \ref{sec:conc}.

\section{Methods}
\label{sec:methods}

\subsection{Stellar Models}
\label{ssec:dartmouth}
The Dartmouth models incorporate both an internal stellar structure
code and a model atmosphere code. Unlike the ATLAS9 models
\citep{castelli+kurucz2004} used in development of the \kepler Input
Catalog, the Dartmouth models perform well for low-mass stars because
the package uses PHOENIX atmospheres to model stars cooler than
10,000K. The PHOENIX models include low-temperature chemistry and are
therefore well-suited for use with low-mass dwarfs
\citep{hauschildt1999a, hauschildt1999b}.

The Dartmouth models include evolutionary tracks and isochrones for a
range of stellar parameters. The tracks and isochrones are available
electronically\footnote{\url{http://stellar.dartmouth.edu/models/grid.html}}
and provide the mass, luminosity, temperature, surface gravity,
metallicity, helium fraction, and $\alpha$-element enrichment at each
evolutionary time step.  We consider the full range of Dartmouth model
metallicities ($-2.5\leq[{\rm Fe}/{\rm H}]\leq0.5)$, but we restrict
our set of models to stars with solar $\alpha$-element enhancement,
masses below $1 \msun$, and temperatures below 7000K. We exclude
models of more massive stars because solar-like stars are well-fit by
the ATLAS9 models used in the construction of the KIC and it is
unlikely that a star as massive as the Sun would have been assigned a
temperature lower than our selection cut $T_{\rm KIC} \leq 5050$K.

The Dartmouth team supplies synthetic photometry for a range of
photometric systems by integrating the spectrum of each star over the
relevant bandpass. We downloaded the synthetic photometry for the 2MASS and Sloan Digital
Sky Survey Systems (SDSS) and used relations~1--4 from
\citet{pinsonneault_et_al2012} to convert the observed KIC magnitudes
for each \kepler target star to the equivalent magnitudes in the SDSS
system. For cool stars, the correction due to the filter differences
is typically much smaller than the assumed errors in the photometry
(0.01 mag in $gri$ and 0.03 mag in $zJHK$, similar to the assumptions
in \citealt{pinsonneault_et_al2012}). All stars have full 2MASS
photometry, but $\photmiss$ of the target stars are missing photometry
in one or more visible KIC bands. For those stars, we correct for the
linear offset in all bands and apply the median correction found for
the whole sample of stars for the color-dependent term. In our final
cool dwarf sample, 70~stars lack $g$-band photometry and 29~stars lack $z$-band. We exclude all stars with more than one missing band. 
 
Our final sample of model stars is drawn from a set of isochrones with
ages 1--13~Gyr and spans a temperature range 2708--6998K. The stars
have masses 0.01--1.00$\msun$, radii 0.102--223$\rsun$, and
metallicities $-2.5 < $~[Fe/H]~$ < 0.5$.  All model stars have solar
$\alpha$/Fe~ratios. There is a deficit of Dartmouth model stars with
radii $0.32 - 0.42\rsun$; we cope with this gap by fitting polynomials
to the relationships between temperature, radius, mass, luminosity,
and colors at fixed age and metallicity. We then interpolate those
relationships over a grid with uniform ($0.01\rsun$) spacing between
0.17$\rsun$ and 0.8$\rsun$ to derive the parameters for stars that
would have fallen in the gap in the original model grid. We compute
the surface gravities for the resulting interpolated models from their
masses and radii. When fitting stars, we use the original grid of
model stars supplemented by the interpolated models. Our fitted parameters 
may be unreliable for stars younger than 0.5~Gyr because 
those stars are still undergoing Kelvin-Helmholtz contraction.

\subsubsection{Distinguishing Dwarfs and Giants}
\label{ssec:giants}
We specifically include giant stars in our model set so that we have
the capability to identify red giants that have been misclassified as
red dwarfs (and vice versa). \citet{muirhead_et_al2012b} discovered
one such masquerading giant (KOI~977) in their spectroscopic analysis
of the cool planet candidate host stars and \citet{mann_et_al2012} have argued that giant
stars comprise $96\% \pm 1\%$ of the population of bright
(Kepmag~$<$~14) and $7\% \pm 3\%$ of the population of dim
(Kepmag~$>$~14) cool target stars. We are confident in the ability of
our photometric analysis to correctly identify the luminosity class of
cool stars because the infrared colors of dwarfs and giant stars are
well-separated at low temperatures. For instance, our photometric
analysis classifies KOI~977 (KIC~11192141) as a cool giant with an
effective temperature of $3894^{+50}_{-54}$K, radius $R_\star =
36^{+3}_{-2}\rsun$, luminosity $L_\star = 260^{+28}_{-25}\lsun$, and
surface gravity $\log g = 1.3^{+0.06}_{-0.05}$. The reported mass
($0.99^{+0.01}_{-0.05}\msun$) is near the edge of our model grid, so
refitting the star with a more massive model grid may yield different
results for the stellar parameters.

\subsection{Revising Stellar Parameters}
\label{ssec:assignments}
We assign revised stellar parameters by comparing the observed optical
and near infrared colors of all 51813~cool ($T_{\rm KIC} \leq 5050$K)
and all 13402~unclassified \kepler planet search target stars to the
colors of model stars.  We account for interstellar reddening by determining the distance at which the apparent $J$-band magnitude of the model star would match
the observed apparent $J$-band magnitude of each target star. We then
apply a band-dependent correction assuming 1 magnitude of
extinction per 1000~pc in $V$-band in the plane of the galaxy
\citep{koppen+vergely1998, brown_et_al2011}.  We find the
best-fit model for each target star by computing the difference in the colors
of a given target star and all of the model
stars. We then scale the differences by the photometric errors in each
band and add them in quadrature to determine the $\chi^2$ for a match
to each model star. 

As explained in Section~\ref{sssec:priors}, we
incorporate priors on the stellar metallicity and the height of stars
above the plane of the galaxy. We rescale the errors so that the
minimum $\chi^2$ is equal to the number of colors (generally 6) minus
the number of fitted parameters (3 for radius, temperature, and
metallicity). We then adopt the stellar parameters corresponding to
the best-fit model and set the error bars to encompass the parameters
of all model stars falling within the $68.3\%$ confidence
interval. For example, for KOI~2626 (KID~11768142), we find $68.3\%$
confidence intervals $R_\star = 0.35\rsun^{+0.11}_{-0.05}$, $T_\star =
3482^{+120}_{-57}$K, and [Fe/H] = $-0.1^{+0.1}_{-0.1}$. We find a
best-fit mass $0.36^{+0.12}_{-0.06}\msun$ and luminosity
$0.016^{+0.02}_{-0.005}\lsun$, resulting in a distance estimate of
$159^{+63}_{-27}$~pc. The corresponding surface gravity is therefore
$\log g = 4.91^{+0.08}_{-0.12}$.

\subsubsection{Priors on Stellar Parameters}
\label{sssec:priors}
We find that fitting the target stars without assuming prior knowledge
of the metallicity distribution leads to an overabundance of
low-metallicity stars, so we adopt priors on the underlying
distributions of metallicity and height above the plane. We then
determine the best-fit model by minimizing the equation
\begin{equation}
 \chi^2_i = \chi^2 _{i, {\rm color}}- 2 \ln P_{{\rm metallicity},i} - 2 \ln P_{{\rm height},i}
\end{equation}
where $\chi^2_{i, {\rm color}}$ is the total color difference between
a target star and model star $i$, $P_{{\rm metallicity},i}$ is the
probability that a star has the metallicity of model star $i$, and
$P_{{\rm height},i}$ is the probability that a star would be found at
the height at which model star $i$ would have the same apparent
$J$-band magnitude as the target star. We weight the priors so that
each prior has the same weight as a single color.

\begin{figure}[htbp]
\begin{center}
\centering
\includegraphics[width=0.5\textwidth]{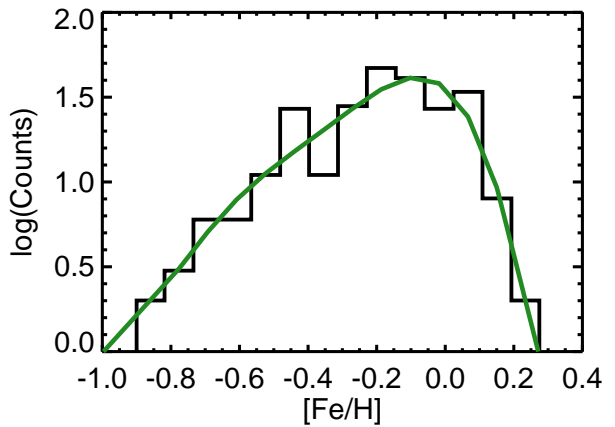}
\end{center}
\caption{Logarithmic number of stars versus logarithmic metallicity
bin. The black histogram displays the distribution of metallicities in
the \citet{casagrande_et_al2008} sample and the green line is our
adopted metallicity prior.}
\label{fig:metpoly}
\end{figure}

We set the metallicity prior by assuming that the metallicity
 distribution of the M~dwarfs in the \kepler target list is similar to
 the metallicity distribution of the 343~nearby M~dwarfs studied by
 \citet{casagrande_et_al2008}. Following \citet{brown_et_al2011}, we
 produce a histogram of the logarithm of the number of stars in each
 logarithmic metallicity bin and then fit a polynomial to the
 distribution. We extrapolate the polynomial down to [Fe/H]~$=-2.5$
 and up to [Fe/H]~$=0.5$ to cover the full range of allowed stellar
 models. Our final metallicity prior and the histogram of M~dwarf
 metallicities from \citet{casagrande_et_al2008} are shown in Figure
 \ref{fig:metpoly}. The distribution peaks at [Fe/H]$=-0.1$ and has a
 long tail extending down toward lower metallicities. We adopt the
 same height prior as \citet{brown_et_al2011}: the number of stars
 falls off exponentially with increasing height above the plane of the
 galaxy and the scale height of the disk is 300~pc
 \citep{cox2000}. Our photometric distance estimates for 77\% of our
 cool dwarfs are within 300~pc, so adopting this prior has little
 effect on the chosen stellar parameters and the resulting planet
 occurrence rate.

\subsection{Assessing Covariance Between Fitted Parameters}
\label{ssec:covar}
Our procedure for estimating stellar parameters expressly considers the covariance between fitted parameters by simultaneously determining the likelihood of each of the models and determining the range of temperatures, metallicities, and radii that would encompass the full 68.3\% confidence interval.  The provided error bars therefore account for the fact that high-metallicity warm M~dwarfs and low-metallicity cool M~dwarfs have similar colors. 

We confirm that the quoted errors on the stellar parameters are large enough to account for the errors in the photometry by conducting  a perturbation analysis in which we create 100~copies of each of the \kepler M~dwarfs and add Gaussian distributed noise to the photometry based on the reported  uncertainty in each band. We then run our stellar parameter determination pipeline and compare the distribution of best-fit parameters for each star to our original estimates. We find that there is a correlation between higher temperatures and higher metallicities, but that our reported error bars are larger than the standard deviation of the best-fit parameters. 

\subsection{Validating Methodology}
\label{ssec:valid}
We confirm that we are able to recover accurate parameters for low mass stars from photometry by running our stellar parameter determination pipeline on a sample of stars with known distances. We obtained a list of 438~M~dwarfs with measured parallaxes, $JHK$ photometry from 2MASS, and $g'r'i'$ photometry from the AAVSO Photometric All-Sky Survey\footnote{\url{http://www.aavso.org/apass}} (APASS) from Jonathan Irwin (personal communication, January 2, 2013) and performed a series of quality cuts on the sample. We removed stars with parallax errors above 5\% and and stars with fewer than two measurements in the APASS database. We then visually inspected the 2MASS photometry of the remaining 230~stars to ensure that none of them belonged to multiple systems that could have been unresolved in APASS and resolved in 2MASS. We removed 203~stars with other stars or quasars within 1', resulting in a final sample of 26~stars. 

\begin{figure}[htbp]
\begin{center}
\centering
\includegraphics[width=0.5\textwidth]{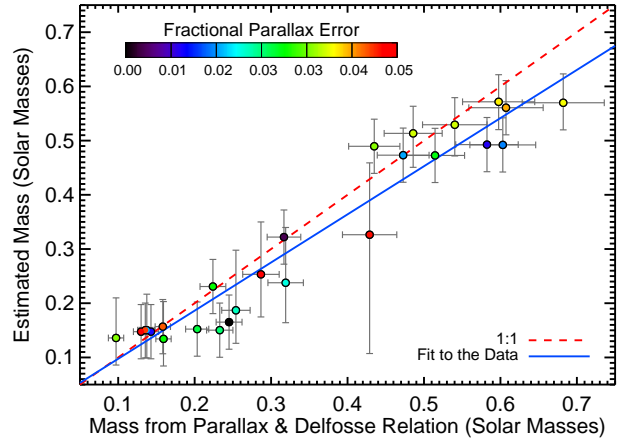}
\end{center}
\caption{Mass estimated by our photometric stellar parameter determination pipeline versus mass predicted by the Delfosse relation. The dashed red line indicates a 1:1 relation and the solid blue line is fit to the data. The points are color-coded by the reported fractional error in the parallax measurement.}
\label{fig:delfosse}
\end{figure}

We estimate the masses of the 26~stars by running our stellar parameter determination pipeline to match the observed colors to the colors of Dartmouth model stars. The APASS $g'r'i'$ photometry was acquired using filters matching the original SDSS $g'r'i'$ bands; we convert the APASS photometry to the unprimed SDSS 2.5m $gri$ bands using the transformation equations provided on the SDSS Photometry White Paper.\footnote{\url{http://www.sdss.org/dr5/algorithms/jeg_photometric_eq_dr1.html}} We then compare the masses assigned by our pipeline to the masses predicted from the empirical relation between mass and absolute $Ks$ magnitude \citep{delfosse_et_al2000}. As shown in Figure~\ref{fig:delfosse}, our mass estimates are consistent with the mass predicted by the Delfosse relation. The masses predicted by the pipeline are typically $5\%$ lower than the mass predicted by the Delfosse relation, but none of these stars have reported $z$-band photometry whereas 96\% of our final sample of \kepler M~dwarfs have full $grizJHK$ photometry. Accordingly, we do not fit for a correction term because the uncertainty introduced by adding a scaling term based on fits made to stars with only five colors would be comparable to the offset between our predicted masses and the masses predicted by the Delfosse relation.

\section{Revised Stellar Properties}
\label{sec:stars}
Our final sample of cool \kepler target stars includes \ndwarfs stars
with temperatures below 4000K and surface gravities above
$\log g = 3.6$. The sample consists primarily of late-K and early-M dwarfs, 
but 201~stars have revised temperatures between $3122-3300$K.
The revised parameters for all of the cool dwarfs are
provided in the Appendix in Table~\ref{tab:stubcoolstars}. We exclude
\ngiants stars from the final sample because their photometry is
consistent with classification as evolved stars ($\log g <3.6$) and \nambig stars because their photometry is insufficient to discriminate between
dwarf and giant models. We refer to the stars that could be fit by either dwarf or giant models as ``ambiguous'' stars. The majority (80\%) of the stars classified as
``ambiguous'' were not assigned temperatures in the KIC.  We find that
$96-98$\% of cool bright ($\teff < 4000$K, Kepmag~$< 14$) stars and
$5-6$\% of cool faint ($\teff < 4000$K, Kepmag~$>14$) stars are
giants, which is consistent with \citet{mann_et_al2012}. (The precise 
fractions of giant stars depend on whether the ambiguous stars 
are counted as giant stars.) One of the
excluded ambiguous stars is KID~8561063 (KOI~961), which was confirmed
by \citet{muirhead_et_al2012a} as a $0.17\pm0.04 \rsun$, $3200 \pm
65$K star hosting sub-Earth-size three planet
candidates. The KIC does not include $z$-band
photometry for KOI~961 and we were unable to rule out matches with
giant stars using only $griJHK$ photometry.

\begin{figure*}[htbp]
\begin{center}
\centering
\includegraphics[width=0.3\textwidth]{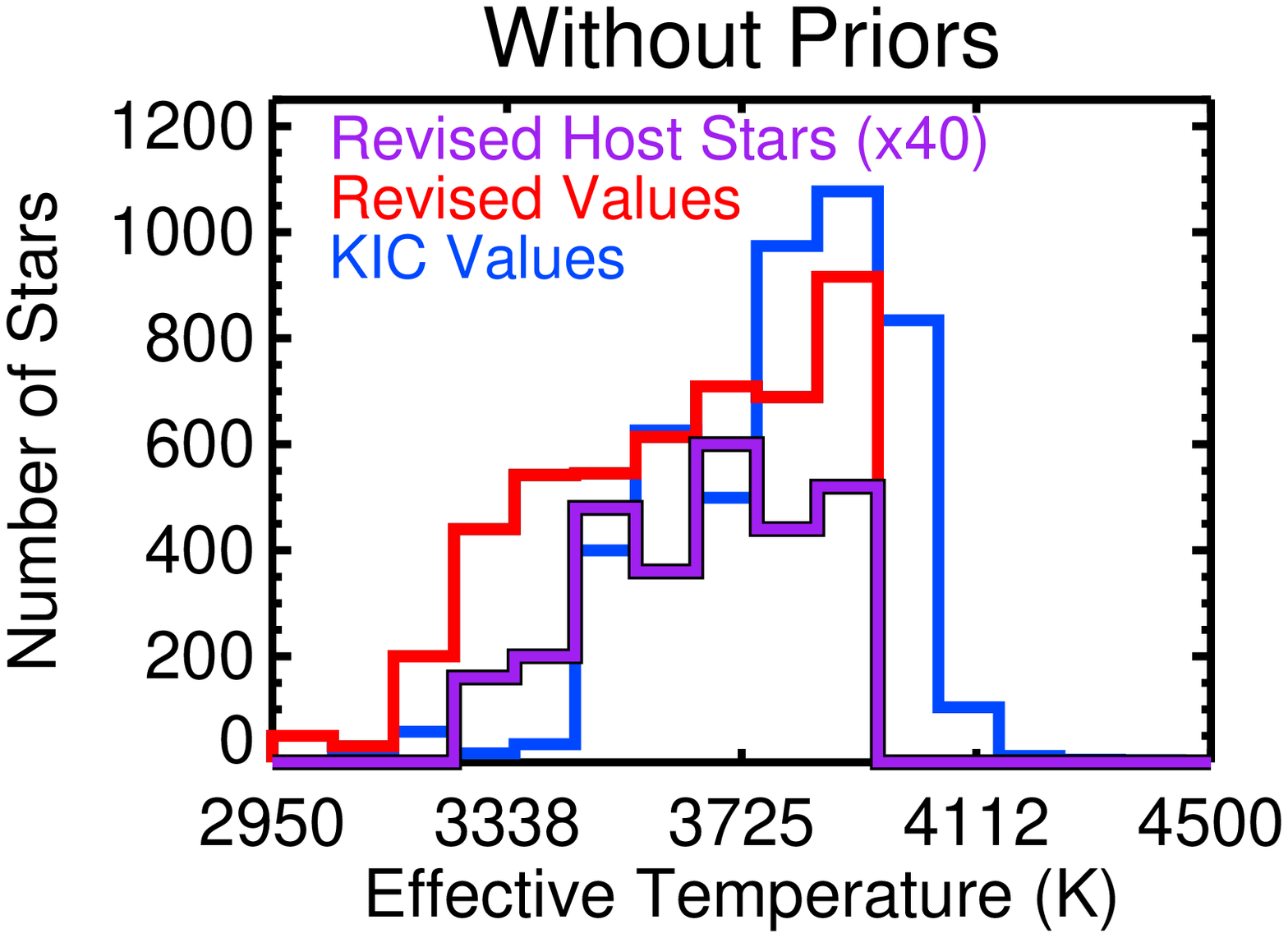}
\includegraphics[width=0.3\textwidth]{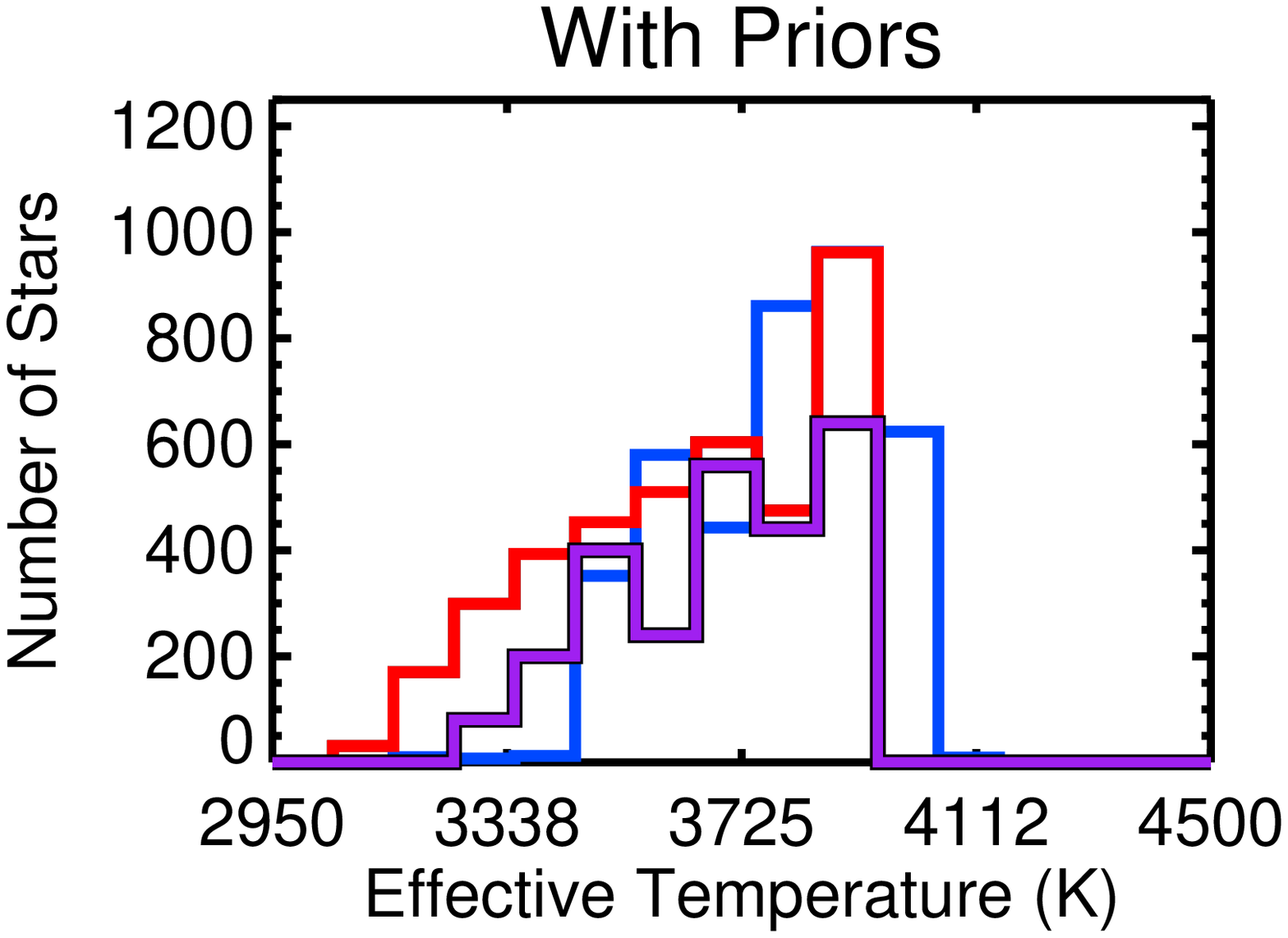} \\
\includegraphics[width=0.3\textwidth]{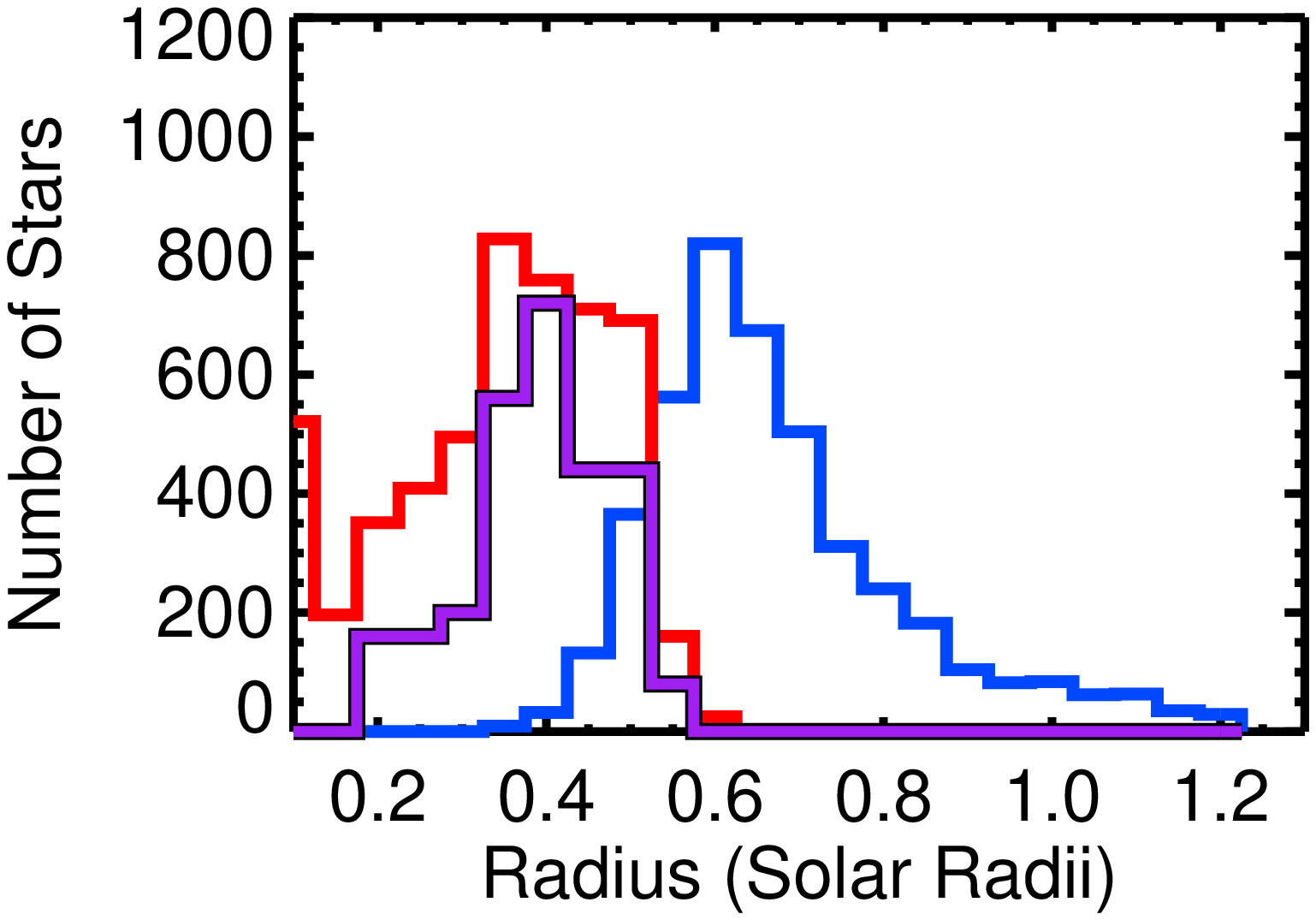}
\includegraphics[width=0.3\textwidth]{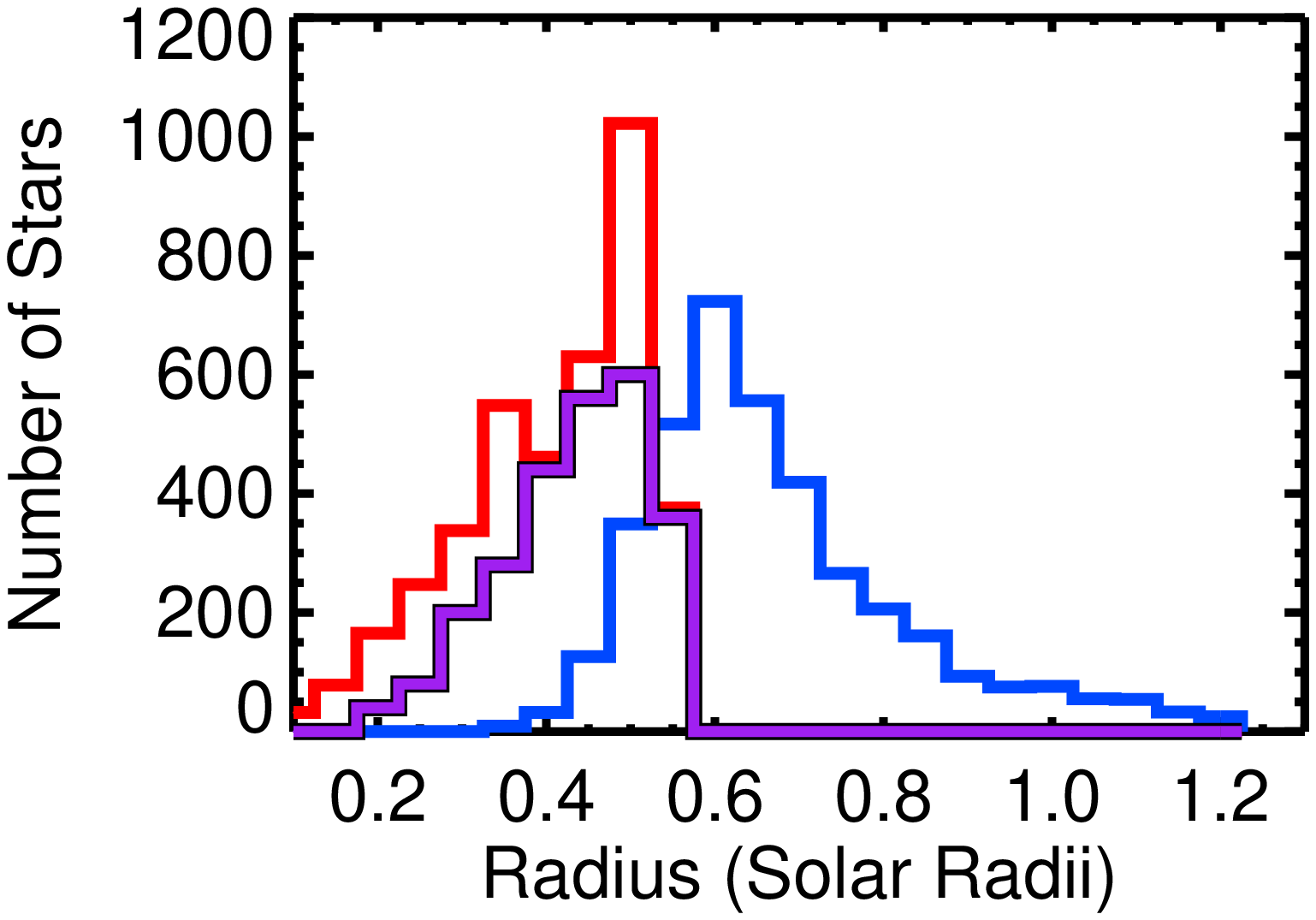} \\
\includegraphics[width=0.3\textwidth]{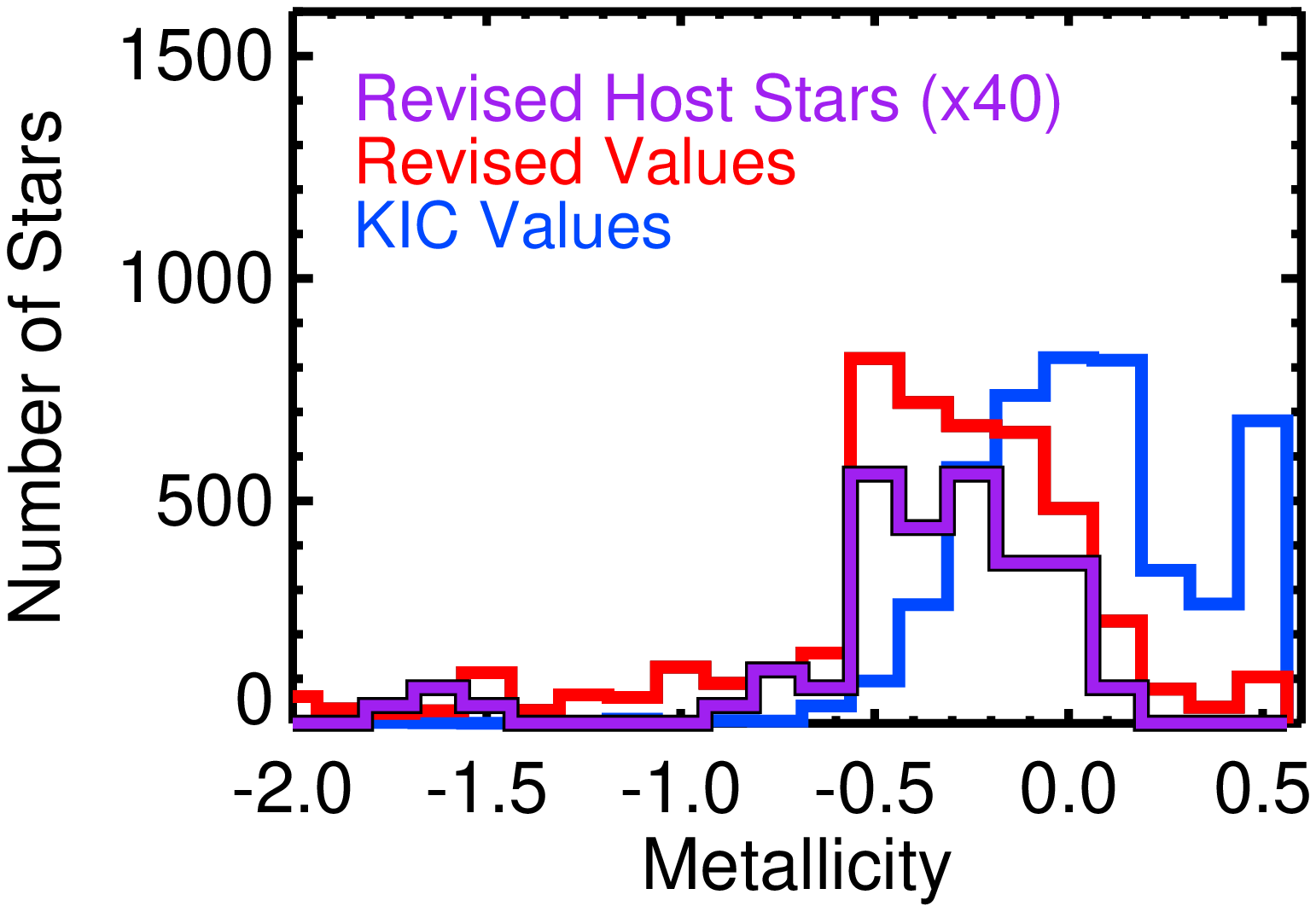}
\includegraphics[width=0.3\textwidth]{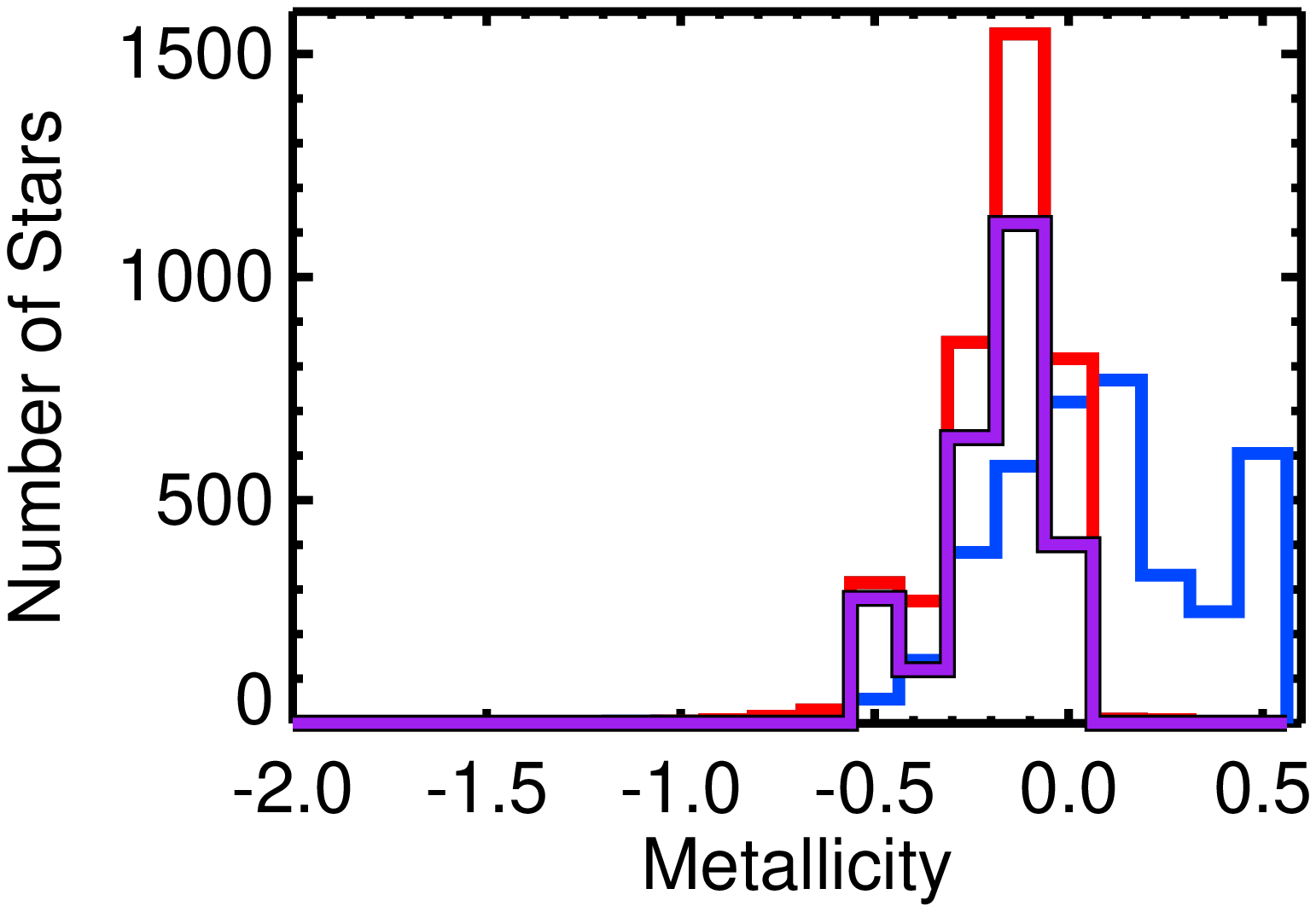} \\
\includegraphics[width=0.3\textwidth]{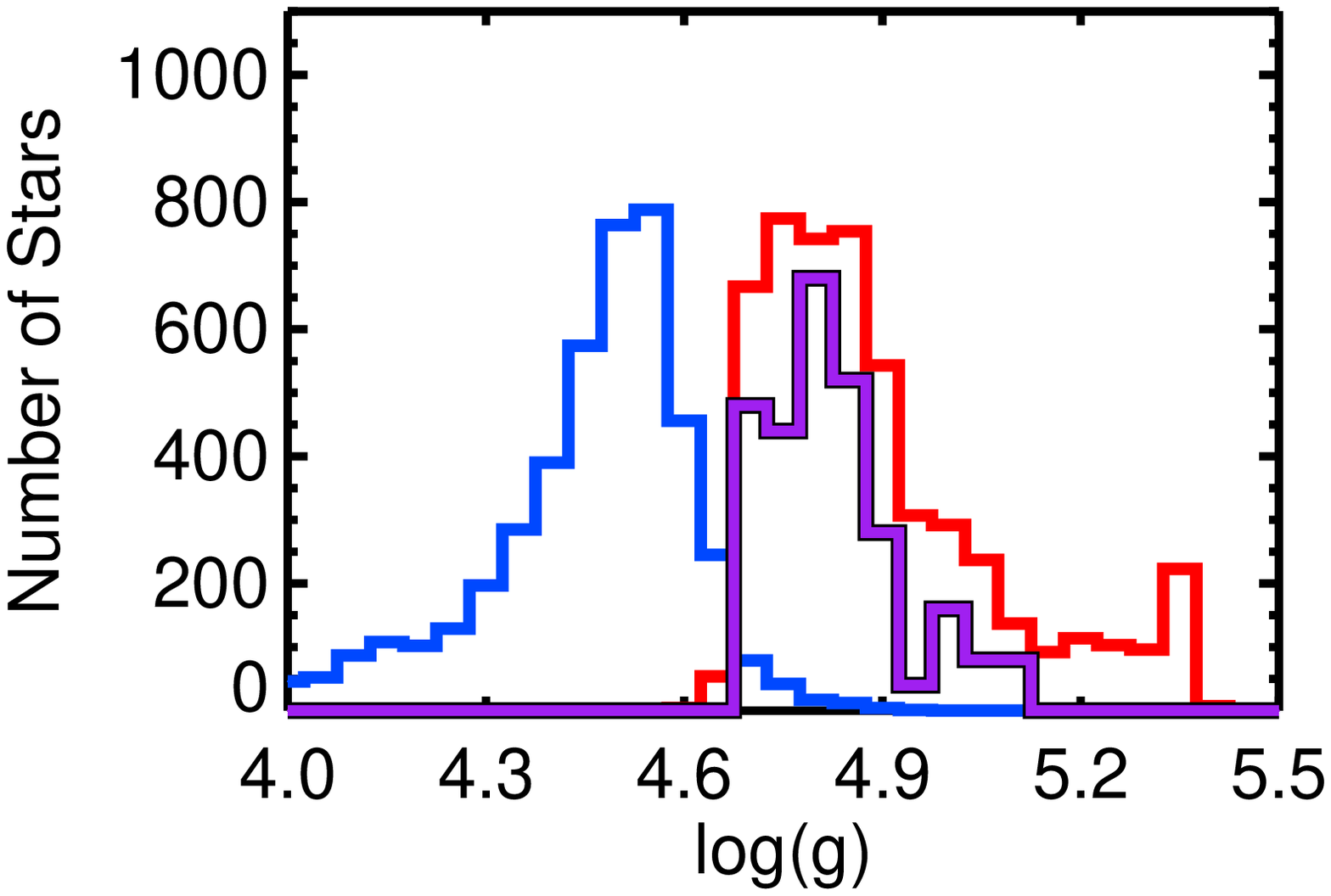}
\includegraphics[width=0.3\textwidth]{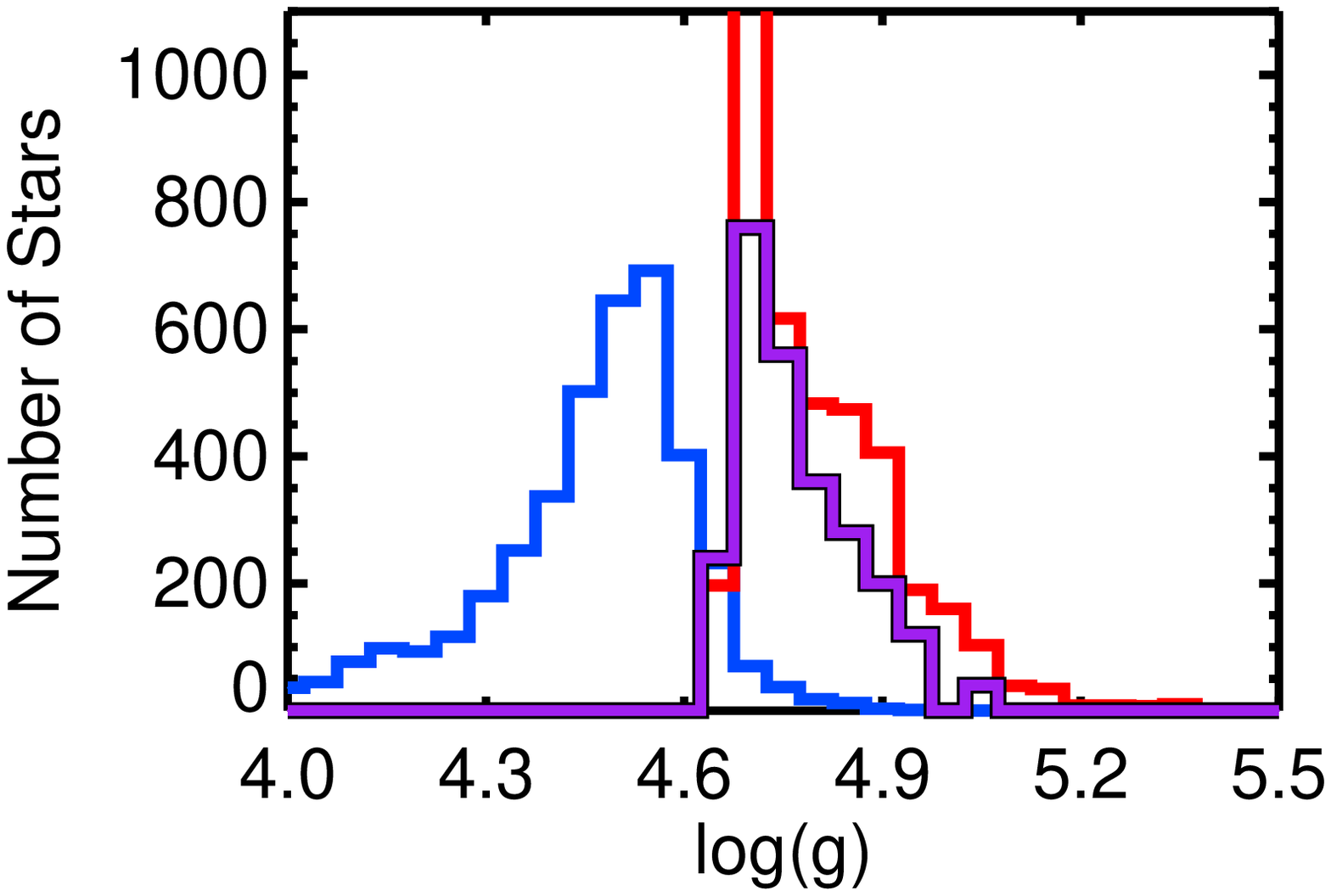} \\
\end{center}
\caption{Histograms of the resulting temperature (top), radius (second
from top), metallicity (third from top), and surface gravity (bottom)
distributions for the target stars with revised temperatures
below~4000K. The panels on the left show the distributions resulting
from fitting the stars without setting priors while the stellar
parameters in the right panels were fit assuming priors on metallicity
and height above the plane. In all panels, a histogram of the original
KIC values is shown in blue and a histogram of the revised values is
plotted in red. The distribution of cool host stars (multiplied by
forty) is shown in purple in all plots.}
\label{fig:stellarfits}
\end{figure*}

The distributions of temperature, radius, metallicity, and surface
gravity for the stars in our sample are shown in Figure
\ref{fig:stellarfits}. For comparison, we display both fits made
without using priors (left panels) and fits including priors on the
stellar metallicity distribution and the height of stars above the
plane of the galaxy (right panels). In both cases the radii of the
majority of stars are significantly smaller than the values given in
the KIC and the surface gravities are much higher.  As discussed in
Section~\ref{sssec:priors}, the primary difference between the two
model fits is that setting a prior on the underlying metallicity
distribution reduces the number of stars with revised metallicities
below [Fe/H]$=-0.6$. Since such stars should be relatively uncommon,
we choose to adopt the stellar parameters given by fitting the stars
assuming priors on metallicity and height above the plane.

Incorporating priors, the median temperature of a star in the sample
is 3723K and the median radius is $0.45\rsun$. Most of the stars in
the sample are slightly less metal-rich than the Sun (median
[Fe/H]=$-0.1$), but 21\% have metallicities
$0.0\leq$[Fe/H]$<0.5$. Although nearly all of the stars in the sample
(96\%) had KIC surface gravities below $\log g = 4.7$, our reanalysis
indicates that 95\% actually have surface gravities above $\log g
=4.7$. As shown by the purple histograms in each of the panels, the
distribution of stellar parameters for the planet candidate host stars matches the
overall distribution of stellar parameters for the cool star sample.

\begin{figure}[htbp]
\begin{center}
\centering
\includegraphics[width=0.5\textwidth]{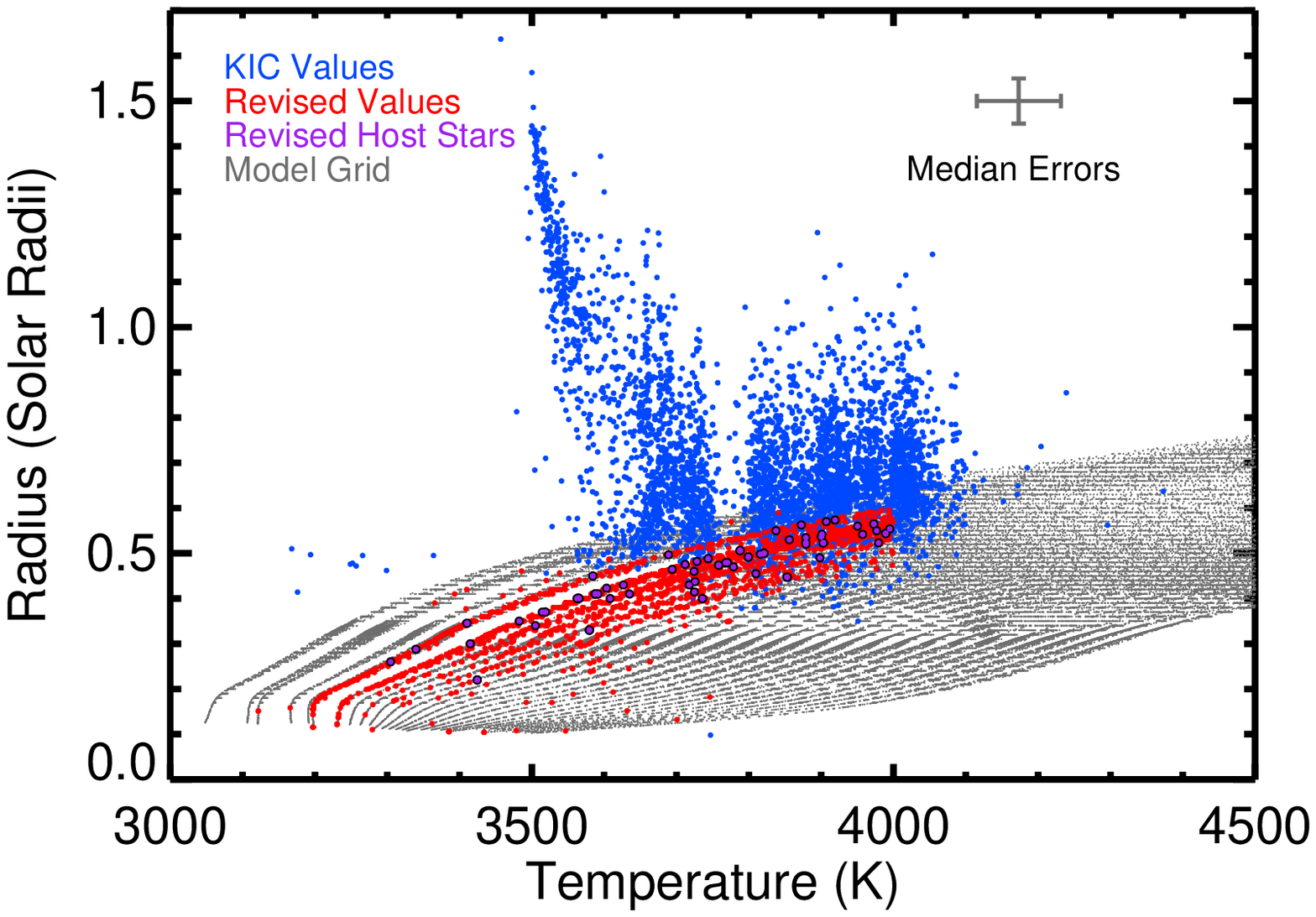}
\end{center}
\caption{Revised (red) and original (blue) temperatures and radii of
the cool target stars. The revised values were determined by comparing
the observed colors of stars to the expected colors of Dartmouth model
stars (gray) and incorporating priors on the metallicity and height
above the galactic plane. The revised stellar parameters for cool planet candidate host
stars are highlighted in purple. The position of the KIC radii well
above the model grid indicates that many of the combinations of radius
and temperature found in the KIC are nonphysical.}
\label{fig:rt}
\end{figure}

The two-dimensional distribution of radii and temperatures for our
chosen model fit is shown in Figure \ref{fig:rt}. The
spread in the radii of the model points at a given temperature is due
to the range of metallicities allowed in the model suite. At a given
temperature, the majority of the original radii from the KIC lie above
the model grid in a region of radius--temperature space unoccupied by
low-mass stars. The discrepancy between the model radii and the KIC
radii is partially due to the errors in the assumed surface
gravities. As shown in Figure~\ref{fig:stellarfits}, the surface
gravities assumed in the KIC peak at $\log(g) = 4.5$ with a long tail
extending to lower surface gravities whereas the minimum expected
surface gravity for cool stars is closer to $\log(g) = 4.7$.

For a typical cool star, we find that the revised radius is only
$69\%$ of the original radius listed in the KIC and that the revised
temperature is 130K cooler than the original temperature estimate. The
majority (96\%) of the stars have revised radii smaller than the radii
listed in the KIC and 98\% of the stars are cooler than their KIC
temperatures. The revised radius and temperature distribution of
planet candidate~host stars is similar to the underlying
distribution of cool target stars. The median changes in radius and
temperature for a cool planet candidate~host star are $-0.19\rsun$ ($-29$\%) and
$-102$K, respectively.

\begin{figure}[htbp]
\begin{center}
\centering
\includegraphics[width=0.5\textwidth]{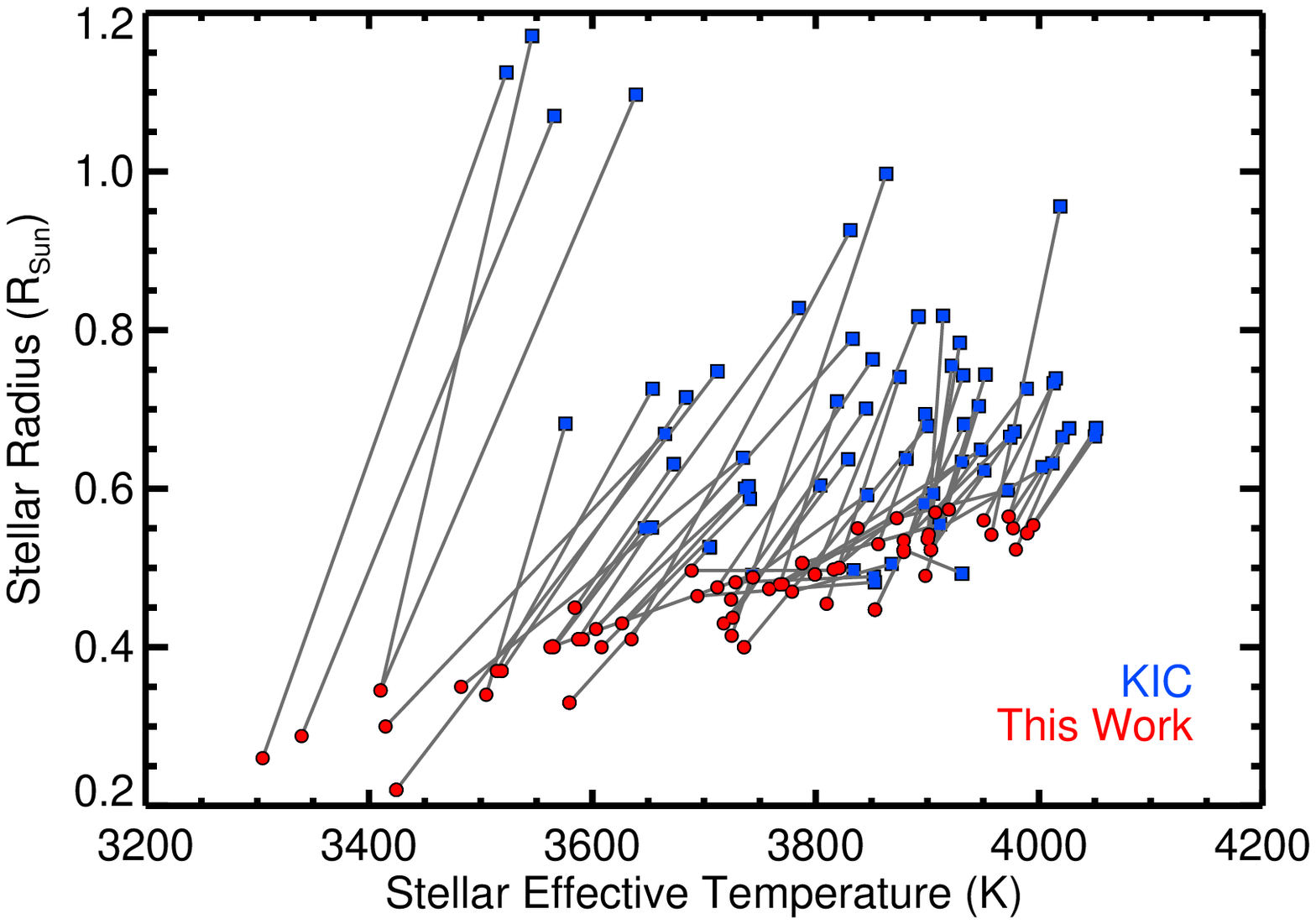}
\end{center}
\caption{Revised (red circles) and original (blue squares) radii and
temperatures for the planet candidate host stars with revised temperatures below
4000K. The gray lines connect the initial and final values for each
host star. }
\label{fig:koirsteff}
\end{figure}

We compare the revised and initial parameters for the host stars in
more detail in Figure~\ref{fig:koirsteff}. For all host stars except for
KOI~1078 (KID~10166274), the revised radii are smaller than 
 the radii listed in the KIC and the
revised temperatures for all of the stars are cooler than the KIC
temperatures. Unlike the original values given in the KIC, the revised
temperatures and radii of the cool stars align to trace out a main
sequence in which smaller stars have cooler temperatures by construction. 

\subsection{Comparison to Previous Work}
\begin{figure}[htbp]
\begin{center}
\centering
\includegraphics[width=0.5\textwidth]{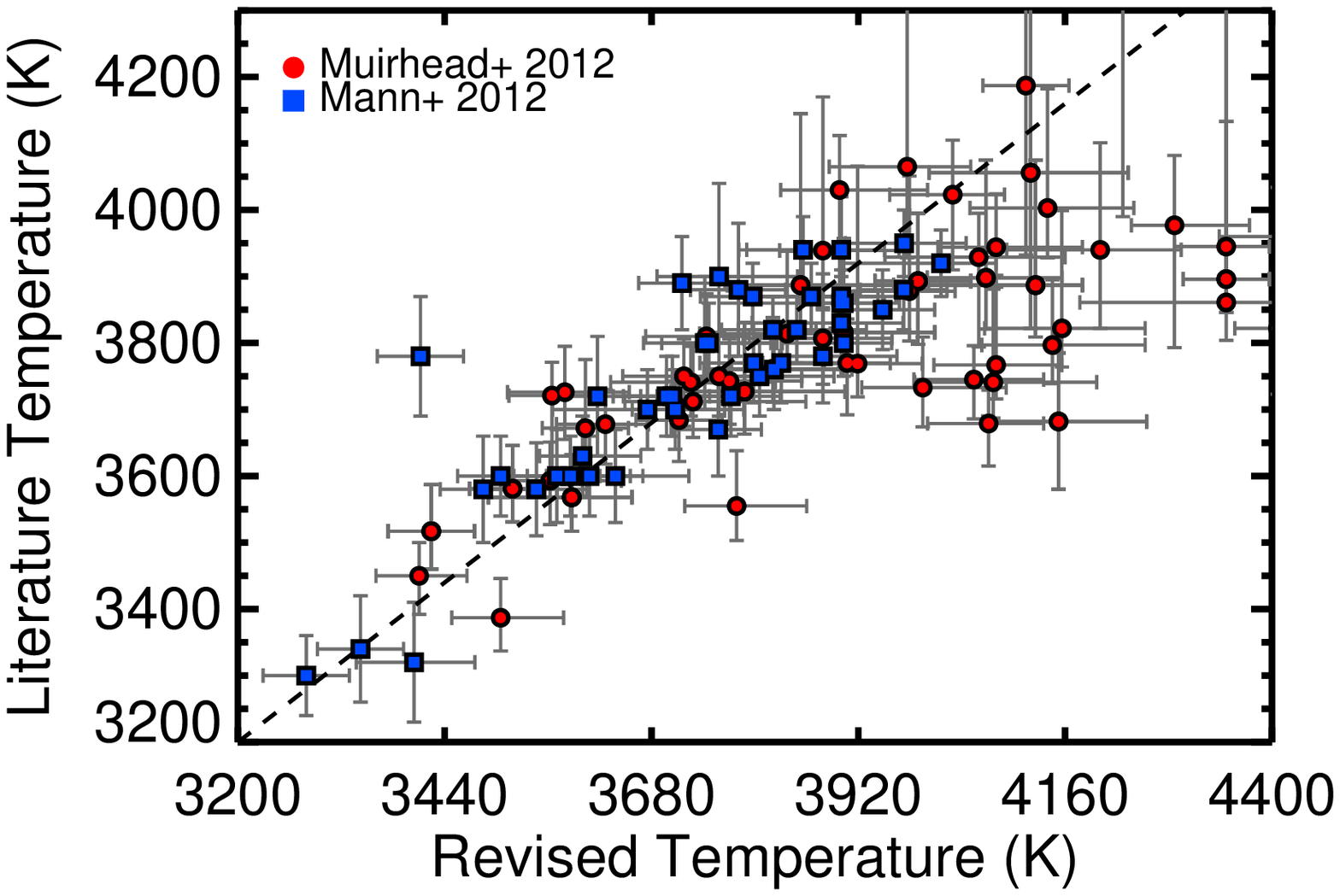}
\end{center}
\caption{Spectroscopic effective temperatures from
\citet{muirhead_et_al2012b} (red circles) and \citet{mann_et_al2012}
(blue squares) versus our revised
photometric effective temperature estimates. The dashed black line indicates a 1:1 relation. The disagreement for the
hotter stars is attributed to the saturation of the H$_2$O-K2 index
used by \citet{muirhead_et_al2012b} at temperatures above 4000K.}
\label{fig:tefflit}
\end{figure}

We validate our revised parameters by comparing our photometric
effective temperatures for a subset of the cool target stars to the
spectroscopic effective temperatures from \citet{muirhead_et_al2012b} and
\citet{mann_et_al2012}. We exclude the stars KIC~5855851 and KIC~8149616 from the comparison due to concerns that their spectra may have been contaminated by light from another star (Andrew Mann, personal communication, January 15, 2013). As shown in Figure~\ref{fig:tefflit}, our revised temperatures are consistent with the literature results for stars with revised temperatures below
4000K, which is the temperature limit for our final sample.

At higher temperatures, we find that our temperatures are
systematically hotter than the literature values reported by
\citet{muirhead_et_al2012b}. The temperatures given in
\citet{muirhead_et_al2012b} are determined from the H$_2$O-K2 index
\citep{rojas-ayala_et_al2012}, which measures the shape of the
spectrum in $K$-band. Although the H$_2$O-K2 index is an excellent
temperature indicator for cool stars, the index saturates
around 4000K, accounting for the disagreement between our temperature
estimates and the \citet{muirhead_et_al2012b} estimates for the hotter
stars in our sample.

\begin{figure}[htbp]
\begin{center}
\centering
\includegraphics[width=0.5\textwidth]{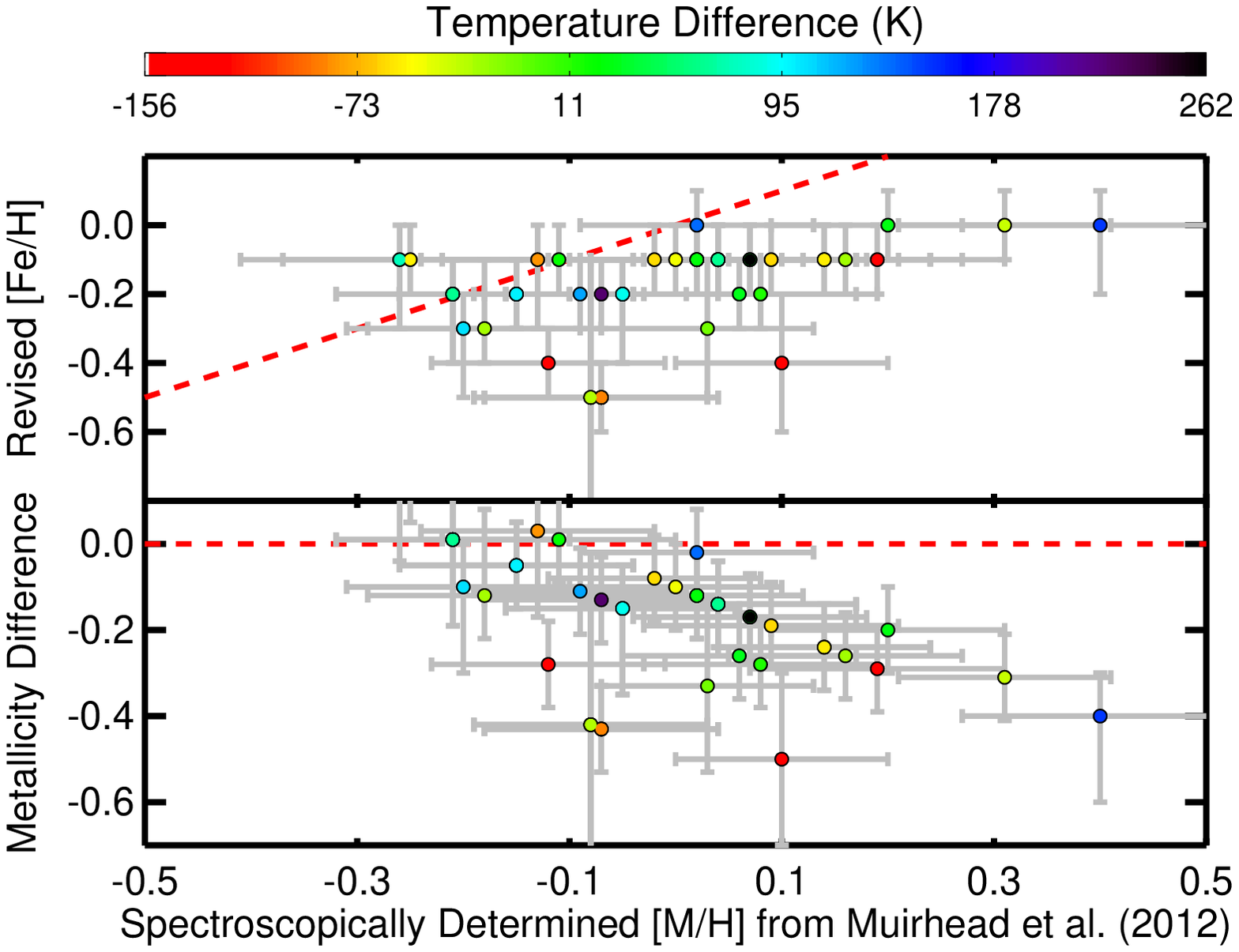}
\end{center}
\caption{Comparison of our photometric metallicity estimates to the
spectroscopic metallicities from \citet{muirhead_et_al2012b} for stars
with revised $T < 4000$K. The color-coding indicates our revised
stellar temperatures and the dashed red lines mark a 1:1 relation between photometric and spectroscopic metallicities. \textbf{Top:} Revised photometric metallicity
estimates versus spectroscopic metallicity. \textbf{Bottom:}
Metallicity difference (photometric - spectroscopic) versus
spectroscopic metallicity.}
\label{fig:metlit}
\end{figure}

We also compare our photometric metallicity estimates to the
spectroscopic metallicity estimates from
\citet{muirhead_et_al2012b}. Given the disagreement between our
temperature estimates and \citet{muirhead_et_al2012b} at higher
temperatures, we choose to plot only the 32~stars with revised
temperatures below 4000K and spectroscopic metallicities from
\citet{muirhead_et_al2012b}. The top panel of Figure~\ref{fig:metlit}
compares our revised metallicities to the spectroscopic metallicities
from \citet{muirhead_et_al2012b}. We observe a systematic offset in
metallicity with our values typically~0.17~dex lower than the
metallicities reported in \citet{muirhead_et_al2012b}.

The metallicity difference is dependent on the spectroscopic metallicity of
the star, as depicted in the lower panel of Figure~\ref{fig:metlit},
which shows the metallicity difference as a function of the
metallicity reported in \citet{muirhead_et_al2012b}. For stars with
\citet{muirhead_et_al2012b} metallicities between -0.2 and -0.1~dex,
our revised metallicities are $0.05$~dex lower, but for stars with
\citet{muirhead_et_al2012b} metallicities above 0.1~dex, our revised
metallicities are 0.3~dex lower.

\section{Revised Planet Candidate Properties}
\label{sec:planets}
Our sample of cool stars includes \nkoi~host stars with
\nplanets~planet candidates. As part of our analysis, we downloaded
the \kepler photometry for the \nplanets~planet candidates and
inspected the agreement between the planet candidate parameters
provided by \citet{batalha_et_al2012} and the \kepler data. We used
long cadence data from Quarters $1-6$ for all KOIs except KOI~531.01,
for which we utilized short cadence data from Quarters~9 and 10 due to
the range of apparent transit depths observed in the long cadence
data. The long cadence data provide measurements of the brightness of the target stars every 29.4~minutes and the short cadence data provide measurements every 58.9~seconds.

\begin{figure}[htbp]
\begin{center}
\centering
\includegraphics[width=0.5\textwidth]{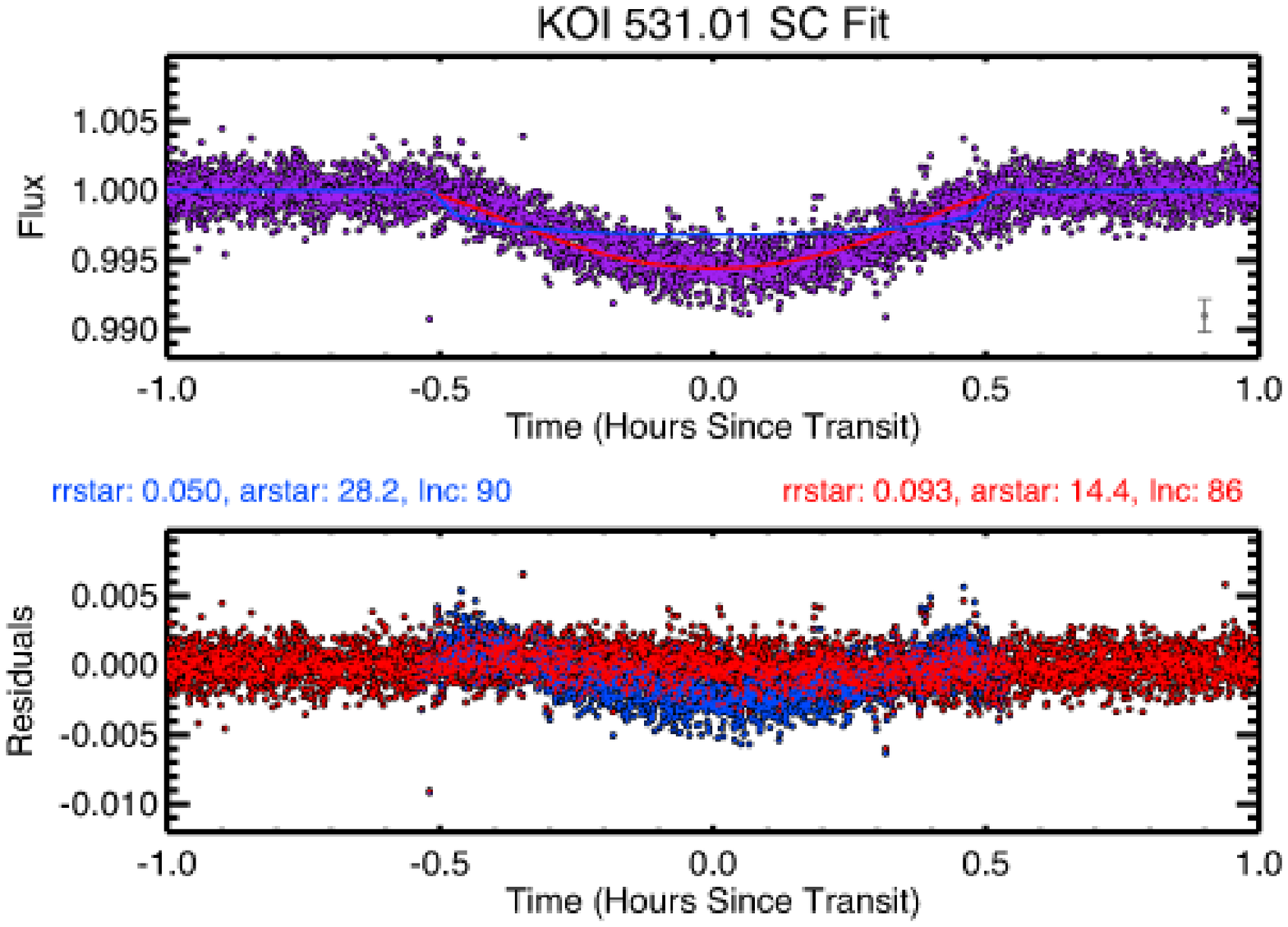}
\end{center}
\caption{Light curve for KOI 531.01. \textbf{Top:} Light curve phased
to best-fit period. The blue curve indicates the original transit model
and the red curve marks our revised fit. For clarity, only 50\% of the 
data are plotted. The gray point in the lower right indicates representative 
error bars. The parameters for the fits
are indicated between the panels. \textbf{Bottom:} Residuals for the
original transit model (blue) and our revised model (red).}
\label{fig:lc53101}
\end{figure}

We detrended the data by dividing each data point by the median 
value of the data points within the surrounding 1000 minute interval 
and masked transits of additional planets in multi-planet
systems. We found that the distribution of impact parameters reported by \citet{batalha_et_al2012} for these planet candidates was biased towards high values (median $b=0.75$) and
that the published parameters for several candidates did not match the observed depth
or shape. Accordingly, we used the IDL AMOEBA minimization
algorithm based on \citet{press_et_al2002} to determine the best-fit
period and ephemeris for each planet candidate. We then ran a Markov Chain Monte
Carlo analysis using \citet{mandel+agol2002} transit models to
revise the planet radius/star radius ratio, stellar radius/semimajor
axis ratio, and inclination for each of the candidates. For each star,
we determined the limb darkening coefficients by interpolating the
quadratic coefficients provided by \citet{claret+bloemen2011} for the
\kepler bandpass at the effective temperature and surface gravity
found in Section \ref{ssec:assignments}. We adopt the median values of
the resulting parameter distributions as our best-fit values and
provide the resulting planet candidate parameters in the Appendix in
Table~\ref{tab:coolkois}. Figures \ref{fig:lc85401}-\ref{fig:lc53101}
display detrended and fitted light curves for the three habitable zone planet
candidates in our sample and for one additional candidate at short cadence. 

\begin{figure}[htbp]
\begin{center}
\centering
\includegraphics[width=0.5\textwidth]{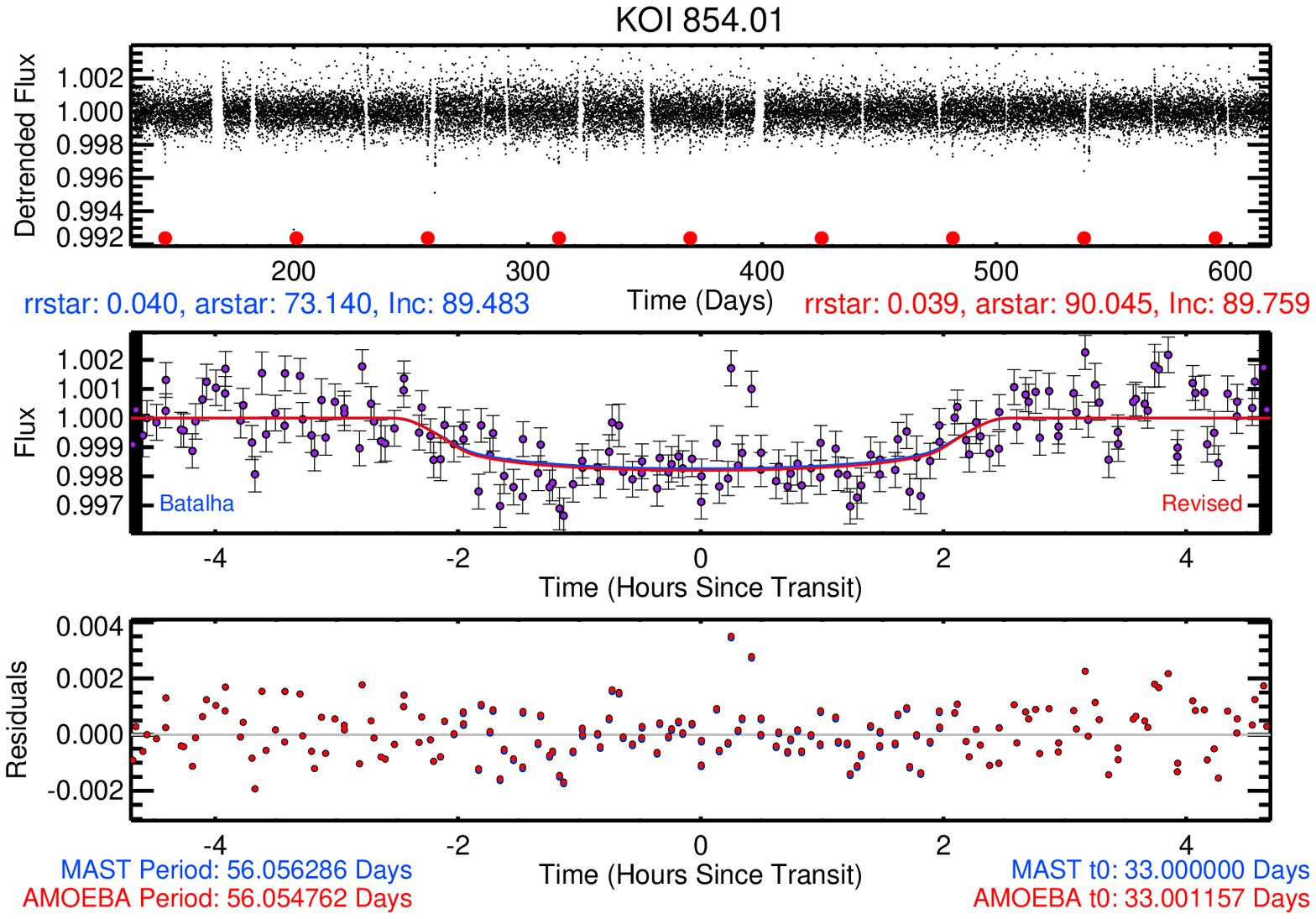}
\end{center}
\caption{Light curve for KOI 854.01. \textbf{Top:} Detrended light
curve with transit times marked by red dots. \textbf{Middle:} Light
curve phased to the best-fit period. The blue curve indicates the
original transit model and the red curve marks our revised fit. The
parameters for the fit are indicated above the middle panel and the
period and ephemeris are marked at the bottom of the figure. The
``MAST'' values indicate the original period and ephemeris listed in
the planet candidate list at MAST and the ``AMOEBA'' values indicate the revised
period and ephemeris. \textbf{Bottom:} Residuals for the original
transit model (blue) and our revised model (red).}
\label{fig:lc85401}
\end{figure}

\begin{figure}[htbp]
\begin{center}
\centering
\includegraphics[width=0.5\textwidth]{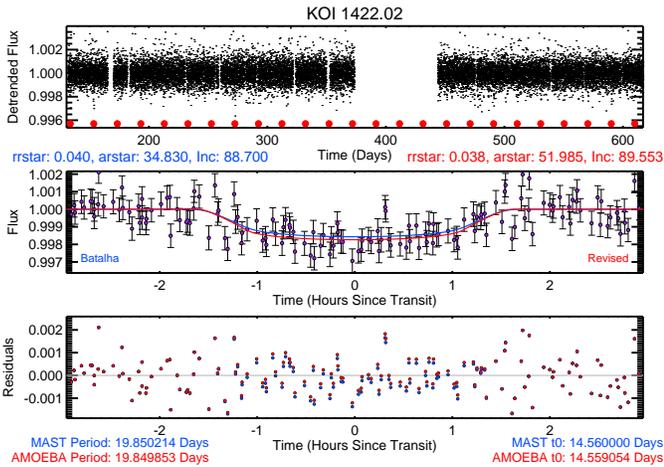}
\end{center}
\caption{Light curve for KOI 1422.02 in the same format as
Figure~\ref{fig:lc85401}.}
\label{fig:lc142202}
\end{figure}

\begin{figure}[htbp]
\begin{center}
\centering
\includegraphics[width=0.5\textwidth]{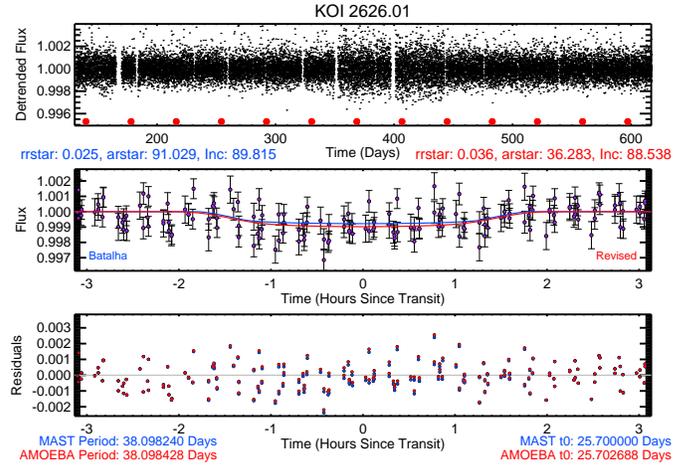}
\end{center}
\caption{Light curve for KOI 2626.01 in the same format as
Figure~\ref{fig:lc85401}.}
\label{fig:lc262601}
\end{figure}

Ten of the planet candidates in our sample have reported transit
timing variations (TTVs), but our fitting procedure assumed a linear
ephemeris. Due to the smearing of ingress and egress caused by fitting
a planet candidate exhibiting TTVs with a linear ephemeris, our simple
fitting routine experienced difficulty determining the transit
parameters for those candidates. Rather than use our poorly
constrained fits for the candidates with TTVs, we choose instead to
adopt the literature values for KOIs~248.01, 248.02, 886.01, and
886.02 (Kepler-49b, 49c, 54b, and 54c) from
\citet{steffen_et_al2012b}, KOIs~250.01 and 250.02 (Kepler-26b and
26c) from \citet{steffen_et_al2012}, KOIs~952.01 and 952.02
(Kepler-32b and 32c) from \citet{fabrycky_et_al2012b}, and KOIs~898.01
and 898.03 from \citet{xie2012}.

We also adopt the transit parameters for KOIs~248.03, 248.04, and
886.03 from \citet{steffen_et_al2012b}, KOI~250.03 from
\citet{steffen_et_al2012}, KOIs~952.03 and 952.04 from
\citet{fabrycky_et_al2012b}, and KOI~254.01 from
\citet{johnson_et_al2012} because the authors completed extensive
modeling of their light curves.  We cannot adopt values from
\citet{steffen_et_al2012} for KOI~250.04 because that planet candidate
was announced after publication of
\citet{steffen_et_al2012}. \citet{fabrycky_et_al2012b} also present
transit parameters for a fifth planet candidate in the KOI~952 system,
but we choose not to add KOI~952.05 to our sample because that planet
candidate was not included in the February 2012 planet candidate list
\citep{batalha_et_al2012} and including KOI~952.05 would necessitate
including any other planet candidates that were not included in the
February 2012 KOI list.

\begin{figure*}[htbp]
\begin{center}
\centering
\includegraphics[width=0.45\textwidth]{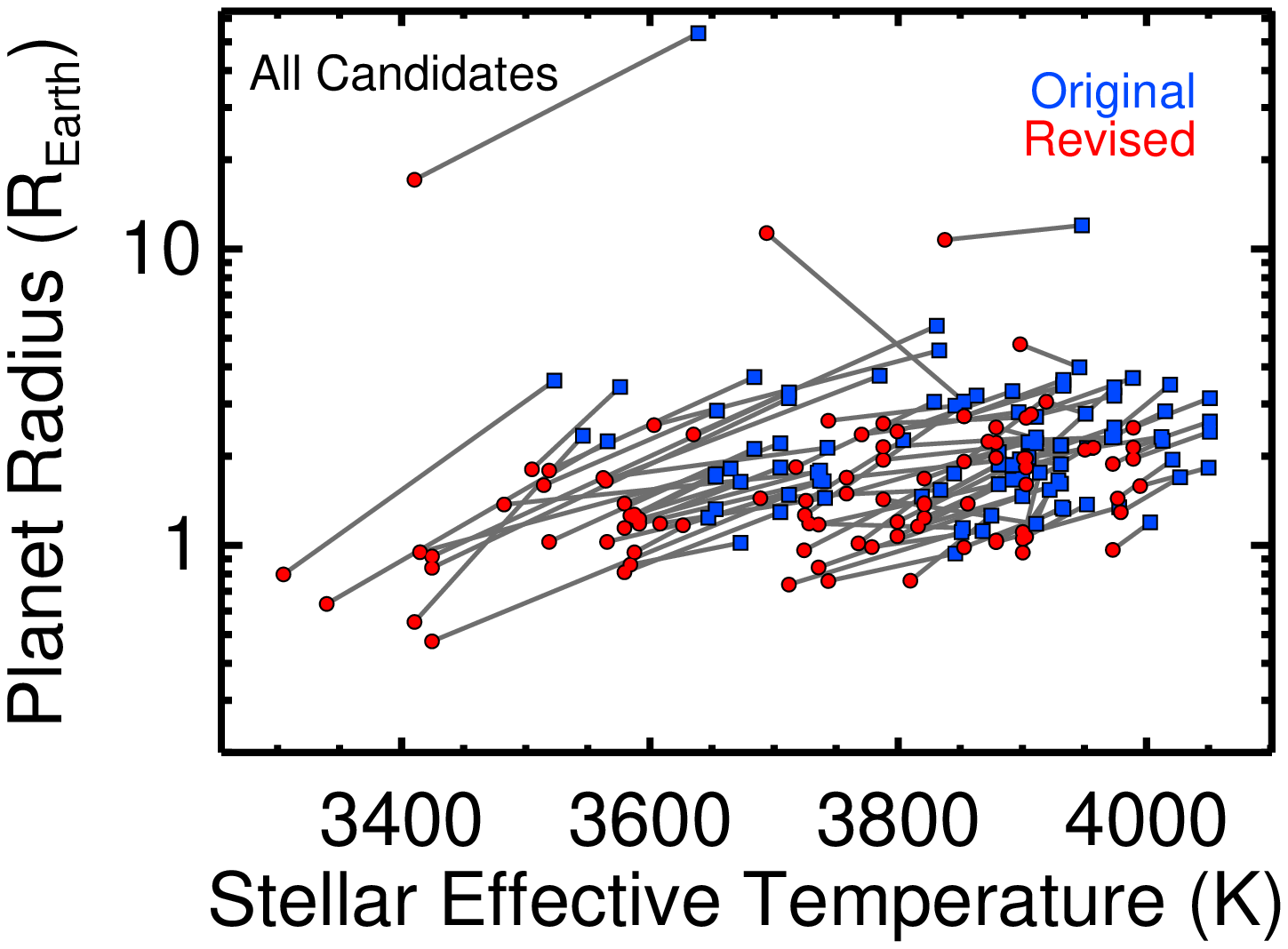}
\includegraphics[width=0.45\textwidth]{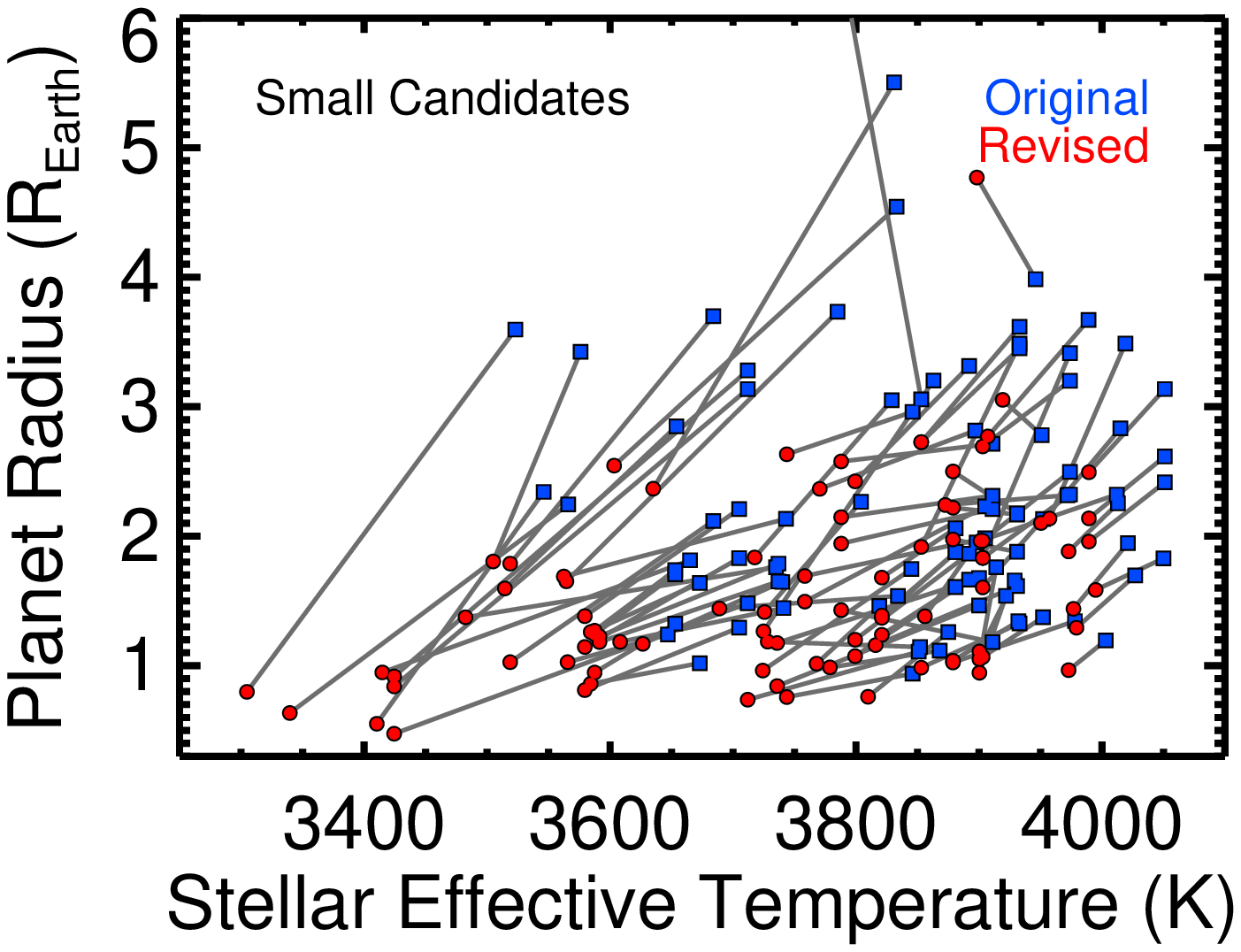}
\end{center}
\caption{Revised (red circles) and original (blue squares) planet
radii and stellar effective temperatures for the \nplanets~planet candidates. The
gray lines connect the initial and final values for each planet
candidate. \textbf{Left:} Full planet candidate
population. \textbf{Right:} Zoomed-in view of the small planet candidate
population.}
\label{fig:rpteff}
\end{figure*}

For 31~of the remaining \nfitplanets~planet candidates without revised
fits from the literature, the planet radius/star radius ratios from
\citet{batalha_et_al2012} lie within the $1\sigma$ error bars of our
revised values. The median changes to the transit parameters for the
refit planet candidates are that the planet radius/star radius ratio
decreases by 3\%, the star radius/semimajor axis ratio increases by
18\%, and the inclination increases by $0.7^\circ$.  Combining our
improved stellar radii with the revised planet radius/star radius
ratios for all of the planet candidates, we find that the radius of a
typical planet candidate is \npshrink~smaller than the value found by
computing the radius from the transit depth given in
\citet{batalha_et_al2012} and the stellar radii listed in the KIC as
shown in Figure~\ref{fig:rpteff}. The improvements in the stellar
radii account for most of the changes in the planet candidate radii,
but the contributions from the revised transit parameters are
non-negligible for a few planet candidates, most notably KOIs~531.01
and 1843.02.

\begin{figure}[htbp]
\begin{center}
\centering
\includegraphics[width=0.5\textwidth]{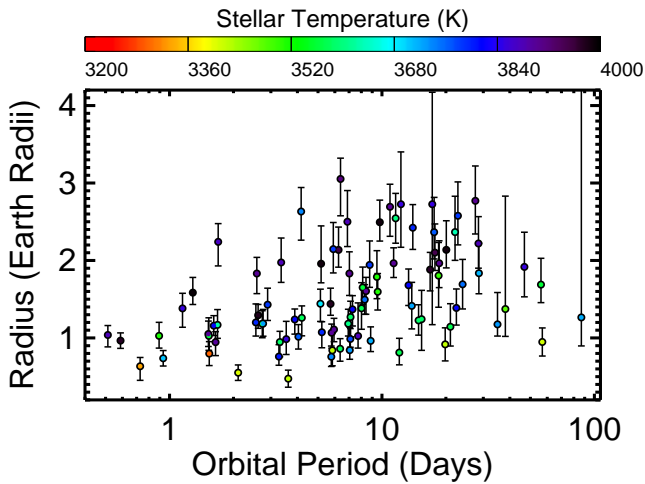}
\end{center}
\caption{Revised planet candidate radius versus orbital period for the
smallest planet candidates. The points are color-coded according to
the temperature of the host star. }
\label{fig:rperr}
\end{figure}

We computed error bars on the planet candidate radii by computing the
fractional error in the planet radius/star radius ratio and the
stellar radius and adding those differences in quadrature to determine
separate upper and lower $1\sigma$ error bounds for each
candidate. For a typical candidate in the sample, the 68\% confidence
region extends from $86-112$\% of the best-fit planet radius. The
best-fit radii and $1\sigma$ error bars for the smallest planet
candidates are plotted in Figure~\ref{fig:rperr} as a function of
orbital period.
 
 \subsection{Multiplicity}
\begin{figure*}[htbp]
\begin{center}
\centering
\includegraphics[width=0.45\textwidth]{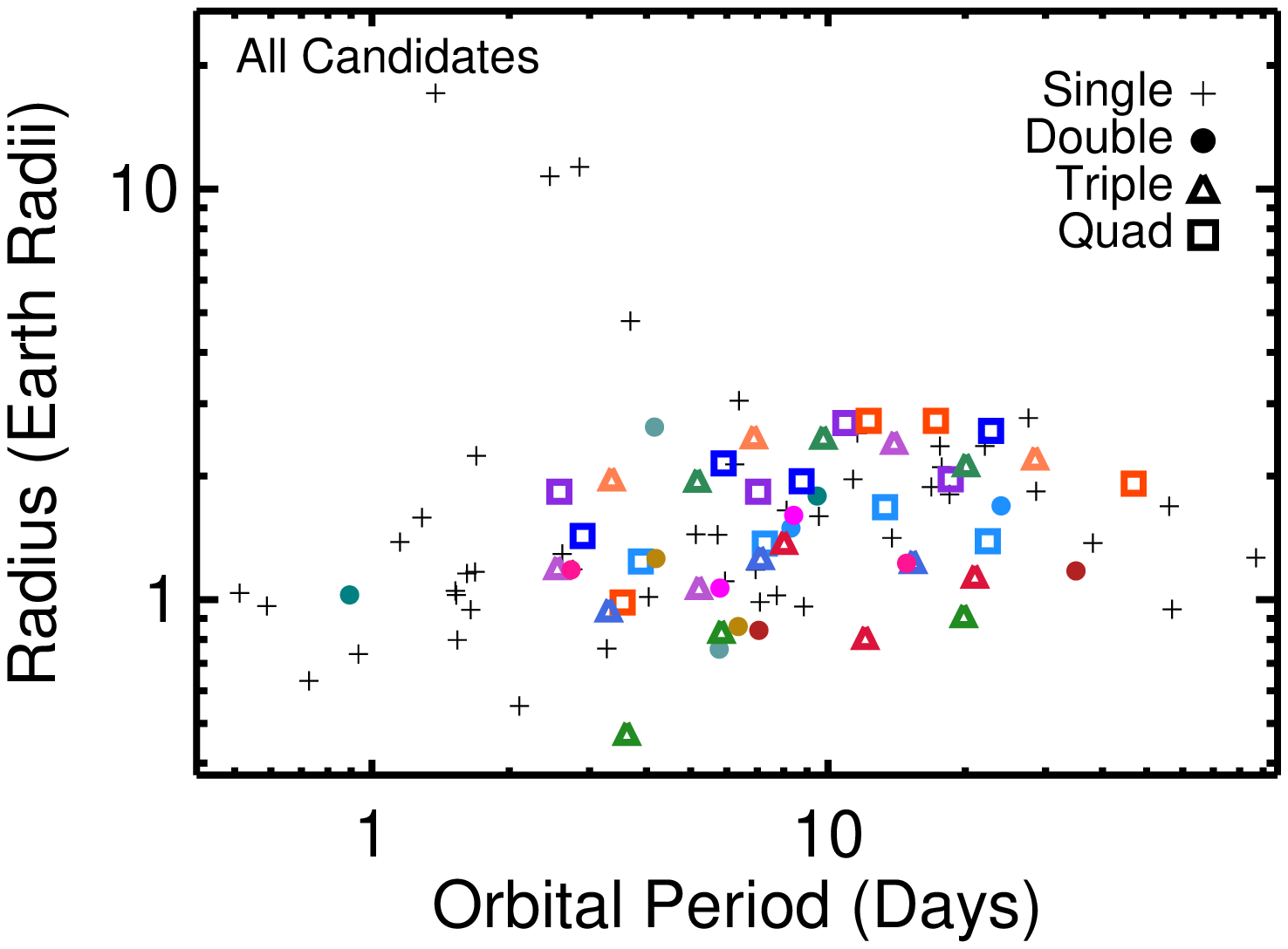}
\includegraphics[width=0.45\textwidth]{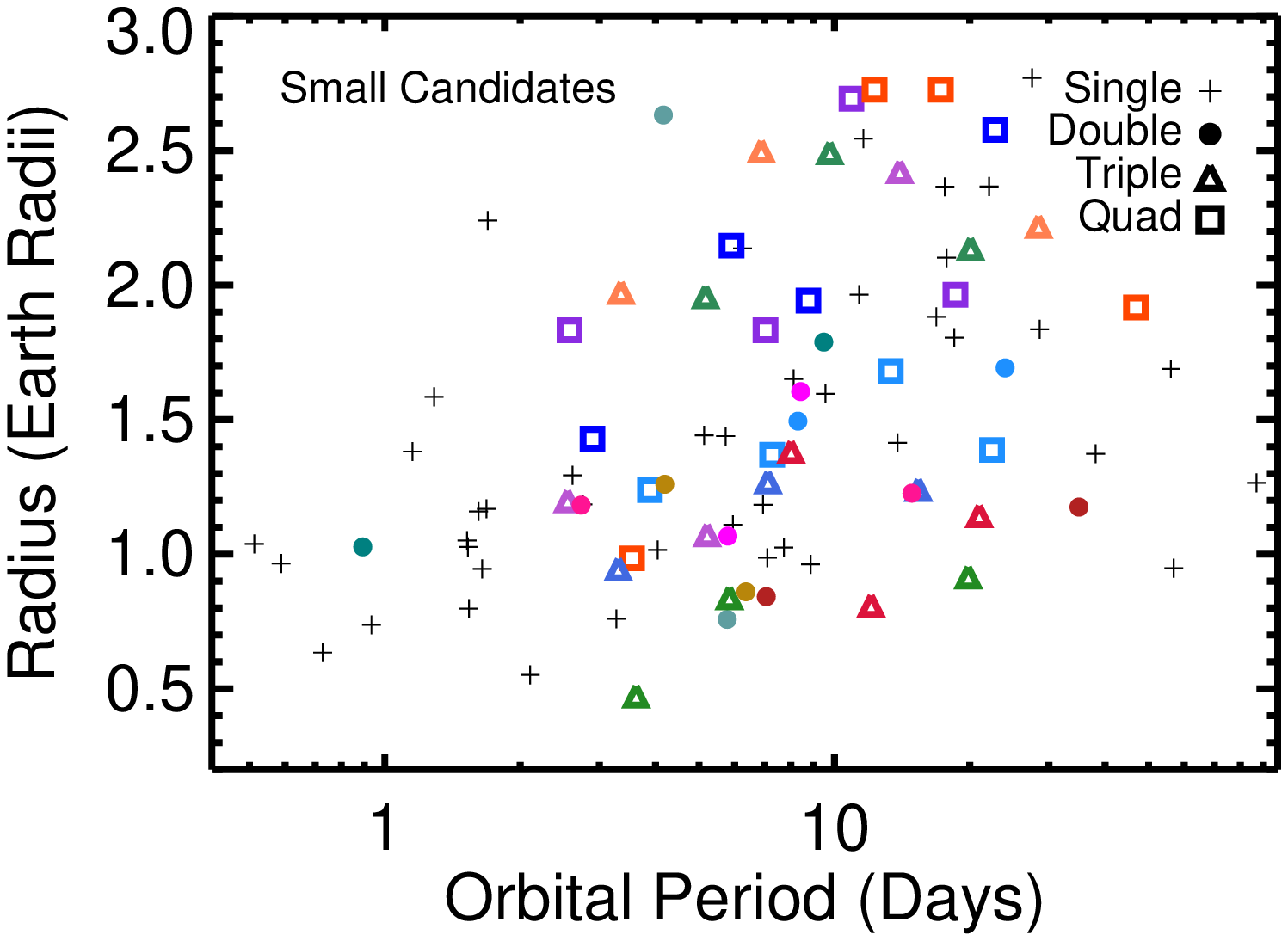}
\end{center}
\caption{Revised planet candidate radii versus orbital period for
candidates in single (cross), double (circle), triple (triangle), and
quadruple (square) systems. Each multi-candidate system is plotted in
a different color. \textbf{Left: }Full planet candidate
population. \textbf{Right: } Zoomed-in view of the smallest planet
candidates.}
\label{fig:multi}
\end{figure*}

Half (48 out of \nplanets) of our cool planet candidates are located in
multi-candidate systems. We mark the multiplicity of each system in
Figure \ref{fig:multi}. As shown in the figure, the largest planet
candidates (KOIs~254.01, 256.01, 531.01, and 2156.01) are in systems
with only one known planet and 93\% of the 14~candidates with orbital
periods shorter than 2 days belong to single-candidate systems. The
one exception is KOI~936.02, which has an orbital period of 0.89~days
and shares the system with KOI~936.01, a $1.8\rearth$ planet in a
9.47~day orbit.  At orbital periods longer than 2~days, 59\% of the
candidates belong to systems with at least one additional planet
candidate. Our sample contains 47~single systems, 7~double systems,
6~triple systems, and 4~quadruple\footnote{\citet{fabrycky_et_al2012b}
report that the KOI~952 system has five planet candidates, but we
count this system as a quadruple planet system because KOI~952.05 was
not included in the February 2012 planet candidate list.}
systems. The fraction of
single planet systems (73\%) is slightly lower than the 79\% single
system fraction for the planet candidates around all stars
\citep{fabrycky_et_al2012a}, but this difference is not significant.

\section{Planet Occurrence Around Small Stars}
\label{sec:occ}
We estimate the planet occurrence rate around small stars by comparing
the number of detected planet candidates with the number of stars
searched. Our analysis assumes that all \nkoi~of the planet candidates are bona fide planet candidates and not false positives. This assumption is reasonable because previous studies have demonstrated that the false positive rate is low for the planet candidates identified by the \kepler team \citep{morton+johnson2011, fressin_et_al2013}.

For a planet with a given radius and orbital period, we
calculate the number of stars searched by determining the depth
$\delta$ and duration of a transit in front of each of the cool
stars. We then calculate the signal-to-noise ratio for a single
transit of each of the stars by comparing the predicted transit depth
to the expected noise level:
\begin{equation}
{\rm SNR}_{\rm 1\, transit} = \frac{\delta}{\sigma_{\rm CDPP}}
\end{equation}
where $\sigma_{\rm CDPP}$ is a measure of the expected noise on the
timescale of the predicted transit duration and the depth $\delta$ for
a central transit is the square of the planet/star radius ratio.

We determine $\sigma_{\rm CDPP}$ by fitting a curve to the observed
Combined Differential Photometric Precision 
(CDPP; \citealt{christiansen_et_al2012}) measured for each
star over 3-hr, 6-hr, and 12-hr time periods and then interpolating to
find the expected CDPP for the predicted transit duration.  CDPP 
is available from the data search form on the \kepler
MAST.\footnote{\url{http://archive.stsci.edu/kepler/data_search/search.php}}

Although the CDPP varies on a quarter-by-quarter basis, we choose to
interpolate the median CDPP value at a given time period for each star
across all quarters. We also repeat our analysis using the minimum and
maximum CDPP for each time interval to quantify the dependence of the
planet occurrence rate on our estimate of the noise in the light curve
on the timescale of a transit.

We then estimate the number of transits $n$ that would have been
observed by dividing the number of days the star was observed by the
orbital period of the planet. We assume that the total signal-to-noise
scales with the number of transits so that the total signal-to-noise
for a planet with radius $R_p$ orbiting a star with radius $R_*$ is:
\begin{equation}
{\rm SNR}_{\rm total} = {\rm SNR}_{\rm 1\, transit} \sqrt{n} = \left(\frac{R_p}{R_*}\right)^2\frac{\sqrt{n}}{\sigma_{\rm CDPP}}
\end{equation}
where $n$ is the number of transits. We adopt the $7.1\sigma$
detection threshold used by the \kepler team and require that the
total SNR is above 7.1$\sigma$ in order for a planet to be
detected. We apply this cut both to our detected sample of planet
candidates and to the sample of stars searched.

\subsection{Correcting for Incomplete Phase Coverage}
Previous research on the occurrence rate of planets around \kepler
target stars has assumed that all stars were observed continuously
during all quarters.  This assumption is reasonable for the objects in the
2011 planet candidate list, but the failure of Module~3 on January 9, 2010
\citep{batalha_et_al2012} means that 20\% of \emph{Kepler's} targets
fall on a failed module every fourth quarter. In addition, some
targets fall in the gaps between the modules and are observed only 1-3
quarters per year even though they never fall on Module~3.

We account for the missing phase coverage by determining the modules
on which each of the stars fall during each quarter and calculating
the fraction of Q1--Q6 that each star spent within the field-of-view of 
the detectors. For a star
that spends $x$ days of the 486.5~day Q1--Q6 observation period 
in the field-of-view of the detectors, we assume 
that $x/486.5$ of transits would be present in the
data. Note that our approach does not account for gaps in phase
coverage during each quarter due to planned events and spacecraft
anomalies. We also ignore the temporal spacing of transits relative to
the gaps in phase coverage. This effect is negligible for transits
that occur multiple times per quarter (i.e., durations $< 90$ days),
but the timing becomes important for transits that occur with periods
equal to or longer than the duration of a quarter.

\subsection{Calculating the Occurrence Rate}
\label{ssec:calcocc}
Following \citet{howard_et_al2012}, we estimate the planet occurrence
rate $f$ as a function of planet radius and orbital period by dividing
the number of planet candidates found with a given radius and period
by the number of stars around which those candidates could have been
detected. We account for non-transiting geometries by multiplying the
number of planet candidates found by the inverse of the geometric
likelihood $p_{\rm transit} = R_*/a$ that a planet with semi-major
axis $a$ would appear to transit a star with radius $R_*$. The planet
occurrence rate over a given period and planet radius range is
therefore:
\begin{equation}
f(R_p, P) = \sum_{i=1}^{N_p(R_p, P)} {\frac{a_i}{R_{*,i}N_{*,i}}}
\label{eqn:occrate}
\end{equation}
where $N_p(R_p, P)$ is the number of planets with the radius $R_p$ and
orbital period $P$ within the desired intervals, $a_i$ is the
semimajor axis of planet $i$ , $R_{*,i}$ is the radius of the host star of
planet $i$, and $N_{*,i}$ is the number of stars around which planet
$i$ could have been detected. Like \citet{howard_et_al2012}, we
estimate the error on the planet occurrence rate $f(R,p)$ by computing
the binomial probability distribution of finding $N_p(R_p, P)$ planets
in a given radius and period range when searching $N_p(R_p, P)/f(R_p,
P)$ stars. We determine the 15.9 and 84.1 percentiles of the
cumulative binomial distribution and adopt those values as the
1$\sigma$ statistical errors on the occurrence rate $f(R_p, P)$ within
the desired radius and period range. 

\begin{deluxetable}{llll}
\tablecolumns{4}
\scriptsize
\tablecaption{Planet Occurrence Rate for Late K and Early M Dwarfs}
\tablehead{
\colhead{} & 
\multicolumn{3}{c}{Orbital Period (Days)}\\
\cline{2-4}\\
\colhead{$R_p (\rearth)$ } &
\colhead{$0.68 - 10$ } &
\colhead{$10-50$} & 
\colhead{$0.68 - 50$}
}
$0.5-0.7$ & $0.014^{+0.0129}_{-0.006}$ (2) & --- & $0.014^{+0.0129}_{-0.006}$ (2)\\[1.2ex]
$0.7-1.0$ & $0.109^{+0.0344}_{-0.025}$ (12) & $0.103^{+0.0977}_{-0.046}$ (2) & $0.212^{+0.0590}_{-0.044}$ (14)\\[1.2ex]
$1.0-1.4$ & $0.108^{+0.0251}_{-0.020}$ (21) & $0.177^{+0.0735}_{-0.048}$ (7) & $0.285^{+0.0509}_{-0.041}$ (28)\\[1.2ex]
$1.4-2.0$ & $0.080^{+0.0245}_{-0.018}$ (13) & $0.123^{+0.0490}_{-0.034}$ (8) & $0.202^{+0.0443}_{-0.035}$ (21)\\[1.2ex]
$2.0-2.8$ & $0.038^{+0.0168}_{-0.011}$ (7) & $0.148^{+0.0456}_{-0.033}$ (12) & $0.186^{+0.0440}_{-0.034}$ (19)\\[1.2ex]
$2.8-4.0$ & $0.005^{+0.0081}_{-0.003}$ (1) & --- & $0.005^{+0.0081}_{-0.003}$ (1)\\[1.2ex]
$4.0-5.7$ & $0.004^{+0.0062}_{-0.002}$ (1) & --- & $0.004^{+0.0062}_{-0.002}$ (1)\\[1.2ex]
$5.7-8.0$ & --- & --- & ---\\[1.2ex]
$8.0-11.3$ & $0.003^{+0.0044}_{-0.001}$ (1) & --- & $0.003^{+0.0044}_{-0.001}$ (1)\\[1.2ex]
$11.3-16.0$ & $0.004^{+0.0055}_{-0.002}$ (1) & --- & $0.004^{+0.0055}_{-0.002}$ (1)\\[1.2ex]
$16.0-22.6$ & $0.003^{+0.0041}_{-0.001}$ (1) & --- & $0.003^{+0.0041}_{-0.001}$ (1)\\[1.2ex]
$22.6-32.0$ & --- & --- & ---\\[1.2ex]
 \enddata
\label{tab:rpocc}
\tablecomments{The number of planets in each bin is given in parentheses.}
\end{deluxetable}

\subsection{Dependence on Planet Size}
\label{ssec:occrp}
\begin{figure*}[htbp] 
\begin{center}
\centering
\includegraphics[width=.9\textwidth]{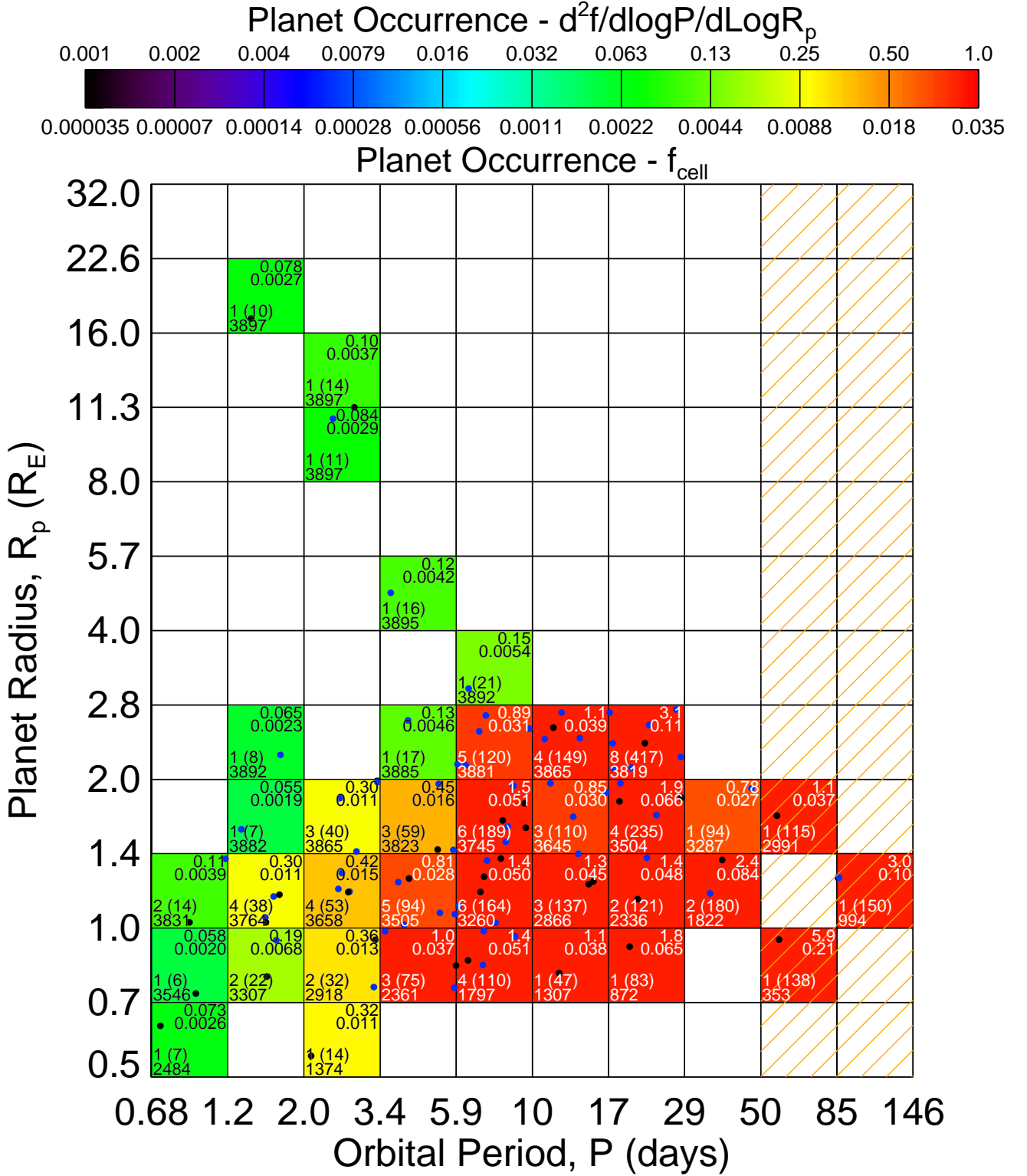}
\end{center}
\caption{Planet occurrence rate as a function of planet radius and orbital period in the style of Figure~4 from \citet{howard_et_al2012}. The color-coding of each cell indicates the planet occurrence within the cell as shown in the legend and the circles mark the radii and periods of the \nplanets~planet candidates in our sample. Planets marked in blue orbit stars hotter than $3723$K and planets marked in black orbit stars cooler than $3723$K. Cells shaded in white do not contain any planet candidates. The planet candidate list is less complete at long periods and our estimates of the planet occurrence rate are likely underestimated at periods longer than 50~days (hatched region). The four numbers within each cell describe the planet occurrence in that region of parameter space: \emph{Top Left:} number of detected planet candidates with signal to noise ratios above $7.1\sigma$ and, in parentheses, the number of non-transiting planets in the same period and radius bin computed by correcting for the geometric probability of transit; \emph{Bottom Left:} the number of stars around which a planet from the center of the grid cell would have been detected with a signal to noise ratio above $7.1\sigma$; \emph{Bottom Right:} the planet occurrence rate within the cell; \emph{Top Right:} planet occurrence per logarithmic area unit.}
\label{fig:howard}
\end{figure*}

Our final sample of planet candidates orbiting dwarf stars with revised
temperatures below 4000K consists of 47~candidates with radii between
$0.5-1.4\rearth$, 43~candidates with radii between $1.4-4\rearth$,
4~candidates with radii above $4\rearth$, and 1~candidate smaller than
$0.5\rearth$. Using Equation~\ref{eqn:occrate}, we find the occurrence
rate of planets with periods shorter than 50~days peaks at
0.29~planets per star for planets with radii between $1.0-1.4\rearth$ and decreases for smaller and larger planets. We summarize our
findings for the occurrence rate as a function of planet radius and
orbital period in Table~\ref{tab:rpocc} and in Figure \ref{fig:howard}. Our estimate for the occurrence rate of planets with radii between $0.5-4\rearth$ and orbital periods shorter than 50~days is \totalseefifty planets per star, which agrees well with the estimate of $1.0^{+0.1}_{-0.1}$ planets per star calculated by \citet{swift_et_al2013}.

We find that the planet occurrence rate per logarithmic bin increases with increasing orbital period and that the occurrence rate of small ($R_{P} < 2.8 \rearth$) candidates with periods less than 50~days is higher than the occurrence rate of larger candidates.  The sample includes only three candidates smaller than $0.7\rearth$, but the low number of planet candidates smaller than $0.7\rearth$ is likely due to incompleteness in the planet candidate list and the inherent difficulty of detecting small planets. In contrast, the scarcity of planet candidates larger than $2.8~\rearth$ indicates that large planets rarely orbit small stars at periods shorter than 50~days.

\begin{figure}[htbp]
\begin{center}
\centering
\includegraphics[width=0.5\textwidth]{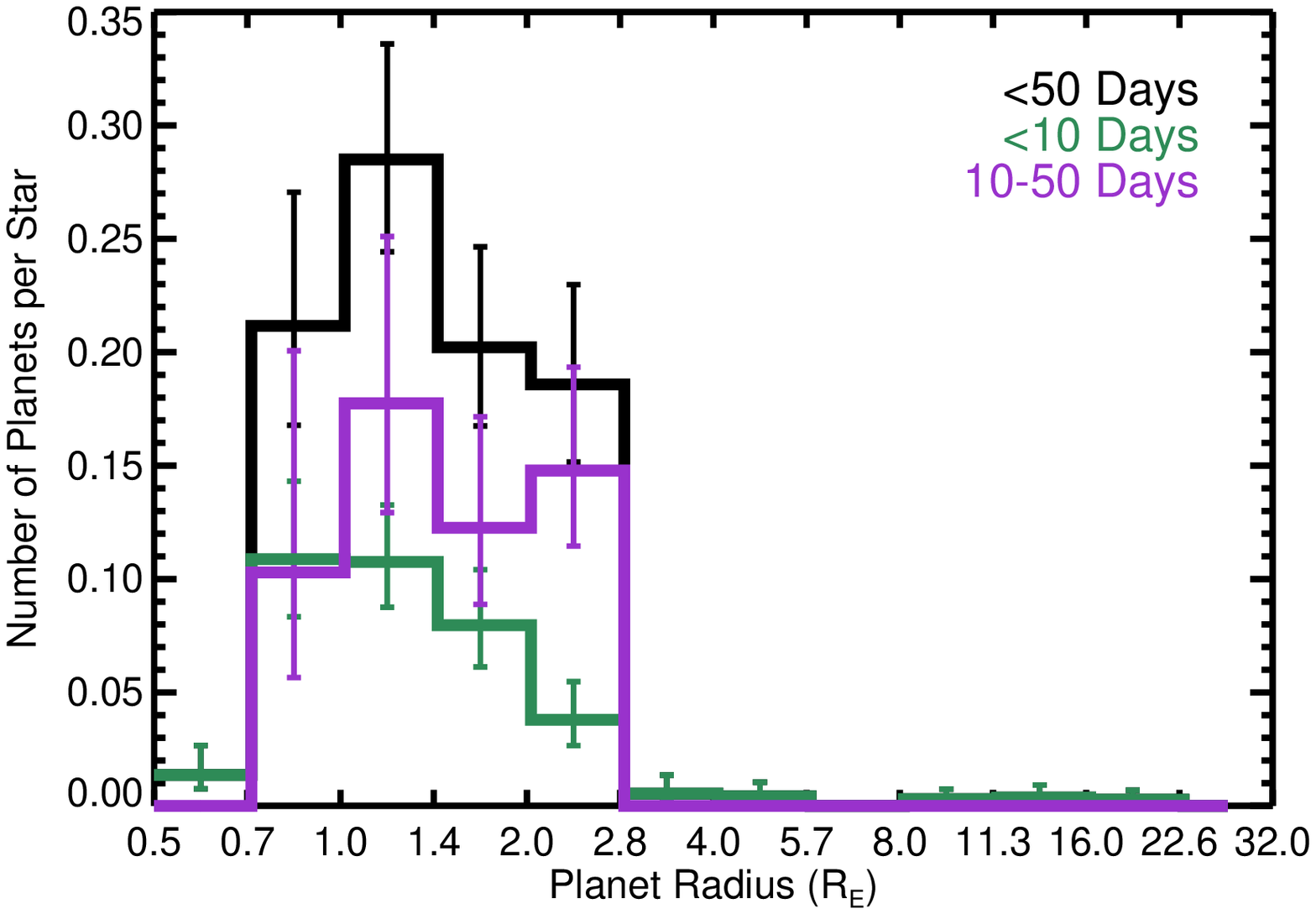}
\end{center}
\caption{Planet occurrence rate as a function of planet radius for all
candidates (black) and candidates with orbital periods shorter than
$<10$ days (green) or between $10-50$ days (purple). The error bars
indicate the errors from binomial statistics and do not include errors
from the stellar and planetary radius estimates.}
\label{fig:rpocc}
\end{figure}

In order to more closely investigate the dependence of the planet occurrence rate on orbital period and planet radius, we plot the occurrence rate as a function of planet radius for planet candidates in three different period groups in Figure~\ref{fig:rpocc}. For the population of candidates with periods shorter than 50~days, we find that the occurrence rate is highest for planets with radii between $1-1.4\rearth$ and decreases at smaller and larger radii. The occurrence rate falls to nearly zero for planets larger than $2.8\rearth$ and to 0.014~planets per star for planets with radii between $0.5-0.7\rearth$. The occurrence rate of planets smaller than $0.7\rearth$ might be underestimated due
to incompleteness in the \kepler pipeline or there might be a real
turnover in the underlying planet radius distribution at small radii.

Breaking down the sample by orbital period, we find a slight indication that 
the planet radius distribution of short-period planets ($P<10$~days) is more
peaked toward smaller planet radii than the distribution of
longer-period planets ($10<P<50$~days), but the difference in the occurrence 
rate is significant only for $2.0-2.8\rearth$ planets. Our result for the 
occurrence rate of $2-4\rearth$ planets within
50~days is 19$^{+5}_{-4}$\%, which is consistent with the
$26^{+8}_{-9}$\% occurrence rate for $2-4\rearth$ planets found by
\citet{howard_et_al2012} for target stars with $3600 \leq T_{\rm eff}
\leq 4100$K. Our result is slightly below the $37\pm 8$\% occurrence
rate for $2-32\rearth$ planets orbiting $3400 \leq T_{\rm eff} \leq
4100$ stars found by \citet{mann_et_al2012} and the 30\% occurrence
rate for $R_{\rm p} \geq 2\rearth$ and $3660 \leq T_{\rm eff} \leq
4660$K found by \citet{gaidos_et_al2012}. \citet{howard_et_al2012}
and \citet{gaidos_et_al2012} adopt the 
KIC parameters for the target stars, so they overestimate both the 
stellar radii and planetary radii 
for the coolest stars in their sample. Accordingly, many of the
planets that we classify as Earth-size would have ended up with radii
above $2\rearth$ in the \citet{howard_et_al2012} and 
\citet{gaidos_et_al2012}, and studies, 
therefore increasing the apparent
occurrence rate of $2-4\rearth$ planets in those studies. 

Additionally, \citet{gaidos_et_al2012}
arrive at their occurrence rate by comparing the number of planet
candidates with radii between $2-32\rearth$ to the number of stars
around which such planets could have been detected, but they use the
noise relation and distribution from \citet{koch_et_al2010} to predict
the expected noise of each star based on \kepler magnitude rather than
using the observed noise. Given that the stellar noise displays
variation even at constant \kepler magnitude, this assumption could
contribute to the slight difference between our occurrence rate and
the value reported by \citet{gaidos_et_al2012}. 

Despite the sensitivity of \kepler to giant planets orbiting small
stars, we find only four planets with {radii $>4\rearth$} in our sample
(KOIs~254.01, 256.01, 531.01, and 2156.01). The implied low occurrence
rate of giant planets is consistent with previous estimates of the
giant planet occurrence rate around cool stars
\citep{butler_et_al2004, bonfils_et_al2006, butler_et_al2006,
endl_et_al2006, johnson_et_al2007, cumming_et_al2008,
bonfils_et_al2011b, howard_et_al2012}. The paucity of giant planets
orbiting M dwarfs is in line with expectations from theoretical
studies of planet formation \citep{laughlin_et_al2004,
adams_et_al2005, ida+lin2005, kennedy+kenyon2008}. The formation of a
giant planet via core accretion requires a considerable amount of
material and the combination of longer orbital timescales and lower
disk surface density decreases the likelihood that a protoplanet will
accrete enough material to become a gas giant before the disk
dissipates.

As an alternative to determining the mean number of planets per star,
we also compute the fraction of stars with planets. The latter number
is more relevant when determining the required number of targets to
survey in a planet finding mission. To compute the fraction of stars
that host planets, we repeat the analysis described in Section
\ref{ssec:calcocc} using only one planet per system. We pick the
planet used for each system by determining which of the planets would
be easiest to detect. We find that $25\%$ of cool dwarfs host planets
with radii $0.5-1.4 \rearth$ and orbital periods shorter than 50~days
and that $25\%$ of cool dwarfs host $1.4-4 \rearth$ planets with
periods shorter than 50~days. These estimates for the fraction of
stars with planets are slightly lower than the mean number of planets
per star due to the prevalence of multiplanet systems.

\subsection{Dependence on Stellar Temperature}
\label{ssec:occteff}
The coolest planet host star (KOI~1702) in our sample has a
temperature of 3305K and the hottest planet host star (KOI~739) has a
temperature of 3995K. The temperature range for the entire small star
sample spans 3122-4000K, with a median temperature of 3723K. Splitting
the cool star population into a cool group ($3122$K$< T_{\rm eff} < \tsplit$K)
and a hot group ($\tsplit$K$ \leq T_{\rm eff} \leq 4000$K), we find that
the cool star group includes 34~KOIs orbiting 25~host stars and the hot
star group includes 61~KOIs orbiting 39~host stars. 
The cool group contains 1957~stars total and the hot group contains 1940~stars total. The multiplicity rates for the two groups are similar: 1.4~planets per
host star for the cooler group and 1.6~planets per host star for the
hotter group.

In order to investigate the dependence of the planet occurrence rate
on host star temperature, we repeat the analysis described in
Section~\ref{sec:occ} for each group separately. We find that the
occurrence rates of Earth-size planets ($0.5-1.4\rearth$) are consistent with
a flat occurrence rate across the temperature range of our sample, but that the occurrence rate of $1.4-4\rearth$ planets is higher for the hot group than for the cool group or for the full sample. The mean numbers of Earth-size planets ($0.5-1.4\rearth$)  and $1.4-4\rearth$
planets per star with periods shorter than 50~days are
$0.57_{-0.06}^{+0.09}$ and $0.61_{-0.06}^{+0.08} $ for the hot group
and $0.46_{-0.06}^{+0.09}$ and $0.19_{-0.05}^{+0.07}$ for the cool
group. 

The lower occurrence rate of $1.4-4\rearth$~planets for the
cool group indicates that cooler M~dwarfs have fewer 
$1.4-4\rearth$~planets than hotter M~dwarfs, but the planet occurrence rate
for mid-M~dwarfs is not well constrained by the \kepler data. Since \kepler is observing 
few mid-M~dwarfs, the median temperature for the 
cool star group is $3520$K and only 26\% of the stars in the 
cool group have temperatures below 3400K. The estimated occurrence rate 
for the cool star group is therefore most indicative of the occurrence rate for stars
with effective temperatures between $3400$K and $\tsplit$K. Further observations 
of a larger sample of M~dwarfs with effective temperatures below 
$3300$K are required to constrain the planet occurrence 
rate around mid- and late-M~dwarfs. 

\subsection{The Habitable Zone}
\label{ssec:hz}
The concept of a ``habitable zone'' within which life could exist is
fraught with complications due to the influence of the spectrum of the
stellar flux and the composition of the planetary atmosphere on the
equilibrium temperature of a planet as well as our complete lack of
knowledge about alien forms of life. Regardless, for this paper we
adopt the conventional and na\"{i}ve assumption that a planet is
within the ``habitable zone'' if liquid water would be stable on the
surface of the planet. For the \nkoi~host stars in our sample, we
determine the position of the liquid water habitable zone by finding
the orbital separation at which the insolation received at the top of
a planet's atmosphere is within the insolation limits determined by
\citet{kasting_et_al1993} for M0~dwarfs. \citet{kasting_et_al1993}
included several choices for the inner and outer boundaries of the
habitable zone. For this paper we adopt the most conservative
assumption that the inner edge of the habitable zone is the distance
at which water loss occurs due to photolysis and hydrogen escape (0.95~AU for the Sun) and the outer edge as
the distance at which CO$_2$ begins to condense (1.37~AU for the Sun).

For M0~dwarfs, these transitions occur when the insolation at the
orbit of the planet is $F_{\rm inner} = 1.00 F_{\oplus}$ and $F_{\rm
outer} = 0.46 F_\oplus $, respectively, where $F_{\oplus}$ is the level of insolation received at the top of the Earth's atmosphere. These insolation levels are
$9\%$ and $13\%$ lower than the insolation at the boundaries of the
G2~dwarf habitable zone because the albedo of a habitable planet is
lower at infrared wavelengths compared to visible wavelengths due to
the wavelength dependence of Rayleigh scattering and the strong water
and CO$_2$ absorption features in the near-infrared. Additionally,
habitable planets around M~dwarfs are more robust against global
snowball events in which the entire surface of the planet becomes covered in ice because increasing the fraction of the planet covered by ice
decreases the albedo of the planet at near-infrared wavelengths and
therefore causes the planet to absorb more radiation, heat up, and
melt the ice. This is not the case for planets orbiting Sun-like stars
because ice is highly reflective at visible wavelengths and because the
stellar radiation peaks in the visible.

We contemplated using the analytic relations derived by
\citet{selsis_et_al2007} for the dependence of the boundaries of the
habitable zone on stellar effective temperature, but the coefficients
for their outer boundary equation were fit to the shape of the maximum
greenhouse limit. The analytic relations derived by \citet{selsis_et_al2007} therefore
 overestimate the position of the edge of the
habitable zone for our chosen limit of the first condensation of
CO$_2$ clouds. Additionally, the equations provided in
\citet{selsis_et_al2007} are valid only for $3700$K$ \leq \teff \leq
7200$K because \citet{kasting_et_al1993} calculated the boundaries of
the habitable zone for stars with temperatures of 3700K, 5700K, and
7200K. \citet{selsis_et_al2007} deals with the lower temperature limit
by assuming that the albedo of a habitable planet orbiting a star with
a temperature below 3700K is sufficiently similar to the albedo of a
habitable planet orbiting a 3700K star that the insolation limits of
the habitable zone are unchanged. In this paper, we extend the
\citet{selsis_et_al2007} approximation to use constant insolation
limits for all of the stars in our sample.  Given the uncertainties
inherent in defining a habitable planet and determining the
temperatures of low-mass stars, our assumption of constant insolation
boundaries should not have a significant effect on our final result
for the occurrence rate of rocky planets in the habitable zones of
M~dwarfs.

\begin{figure*}[htbp]
\begin{center}
\centering
\includegraphics[width=0.45\textwidth]{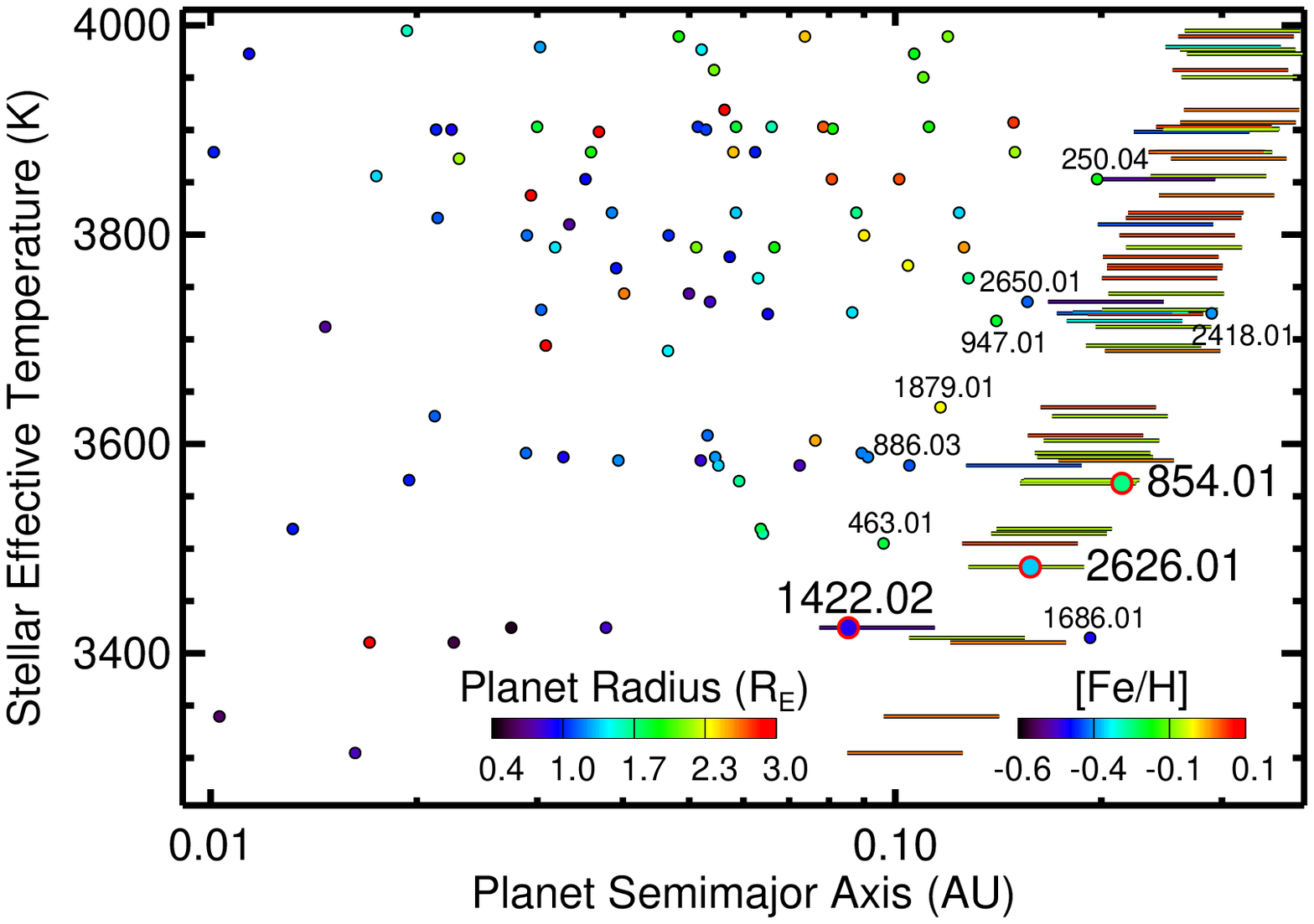}
\includegraphics[width=0.45\textwidth]{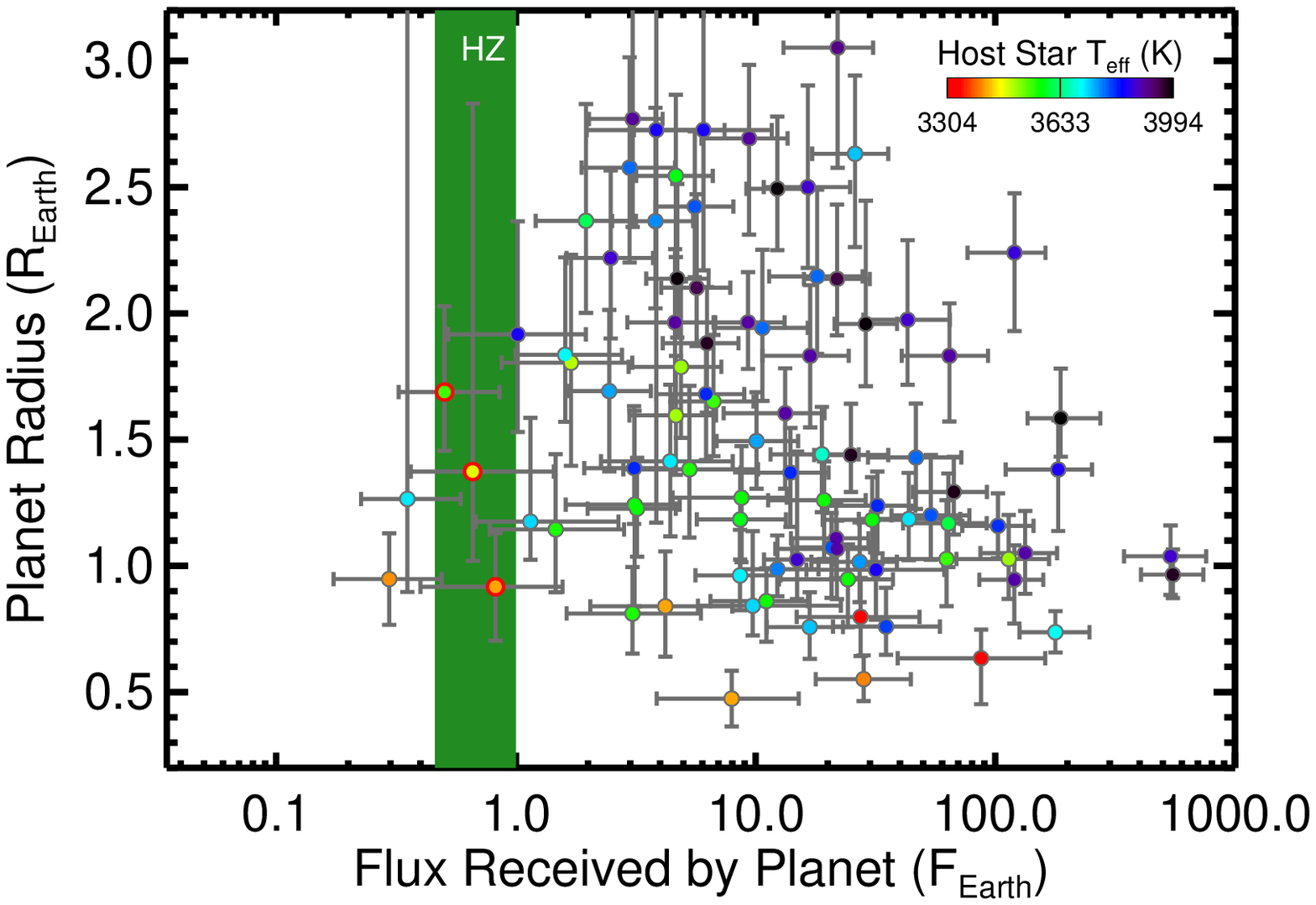}
\end{center}
\caption{\textbf{Left: }Stellar effective temperature and planet
semimajor axes for the \nplanets~planet candidates orbiting stars with revised
temperatures below 4000K. The points are color-coded according to the
radius of each planet candidate as indicated in the left legend. The
lines indicate the calculated position of the habitable zone (HZ) for
each star and are color-coded accorded to the metallicity of the star
as indicated in the right legend. The three candidates within the HZ
(KOIs~854.01, 1422.02, and 2626.01) are identified by name and
highlighted in red. \textbf{Right: }Planet radii versus flux for the
planet candidates around stars with revised temperatures below 4000K. The
color-coding indicates the effective temperature of the host star. The
green box indicates the habitable zone as defined
in Section \ref{ssec:hz}.}
\label{fig:hz}
\end{figure*}

\subsection{Planet Candidates in the Habitable Zone}
As shown in Figure \ref{fig:hz}, the habitable zones for the
\nkoi~host stars in our final sample of dwarfs cooler than 4000K fall
between 0.08 and 0.4~AU, corresponding to orbital periods of
$17-148$~days. Figure~\ref{fig:hz} displays the semimajor axes of all
of the planet candidates and the positions of the habitable zones
around their host stars. Nearly all of the planet candidates orbit
closer to their host stars than the inner boundary of the habitable zone, but
two candidates (KOIs 1686.01 and 2418.01) orbit beyond the habitable
zone and two candidates (KOI~250.04 and 2650.01) orbit just inside the
inner edge of the habitable zone. Three candidates fall within our
adopted limits: KOIs~854.01, 1422.02, and 2626.01. These candidates
are identified by name in Figure \ref{fig:hz} and have radii of 1.69,
0.92, and 1.37$\rearth$, respectively. A full list of the stellar and
planetary parameters for the three candidates in the habitable zone
and the candidates near the habitable zone is provided in
Table~\ref{tab:hz}.

\begin{figure}[htbp]
\begin{center}
\centering
\includegraphics[width=0.5\textwidth]{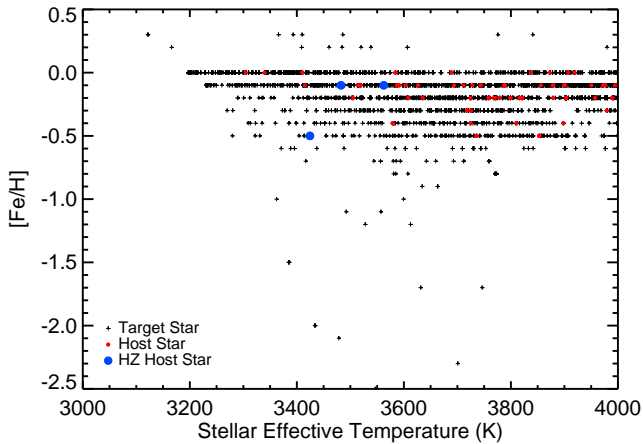}
\end{center}
\caption{Revised metallicities versus stellar effective temperature
for all stars with revised temperatures below 4000K (black crosses)
and planet candidate host stars (circles). The three stars hosting planet candidates within the
habitable zone are highlighted in blue; all other planet host stars
are marked in red.}
\label{fig:teffmet}
\end{figure}

The lateral variation in the position of the habitable zone at a given
stellar effective temperature is due to the range of metallicities
found for the host stars. At a given stellar effective temperature,
stars with lower metallicities are less luminous and therefore the
habitable zone is located closer to the star. Adopting a different
metallicity prior would change the metallicities of the host stars and
shift the habitable zones slightly inward or outward. The
metallicities and temperatures of the cool stars and planet candidate host stars
are plotted in Figure \ref{fig:teffmet}. As shown in Figure
\ref{fig:teffmet}, 98\% of the cool stars and all of the planet candidate host
stars have metallicities $-0.5 \leq [$Fe/H$] \leq 0$. There are
17~cool stars (0.4\%) with super-solar metallicities and 75~cool
stars (2\%) with metallicities below [Fe/H]$=-0.5$.

All of the habitable zone candidates orbit stars fit by models with
sub-solar metallicity (KOI~854: [Fe/H]$=-0.1$, KOI~1422:
[Fe/H]$=-0.5$, KOI~2626: [Fe/H]$=-0.1$). If we restrict all of the
stars to solar metallicity and redetermine the stellar parameters and
habitable zone boundaries for each planet candidate, then we find that the number
of candidates in the habitable zone remains constant, but that
identity of the habitable zone candidates changes. KOIs~854.01 and
2626.01 remain in the habitable zone, but KOI~1422.02 does not. We
find that the habitable zones of KOIs~1422 and 2418 move outward so
that KOI~1422.02 is now too close to the star to be within the
habitable zone and that KOI 2418.01 is now within the boundaries of
the habitable zone. Because the number of candidates in the habitable 
zone is unchanged, our estimate of the occurrence rate within the 
habitable zone is not affected by adopting a different metallicity prior.

\begin{deluxetable*}{rrllllll}
\tablecolumns{7}
\tablewidth{0pt}
\tablecaption{Properties of Candidates In or Near the Habitable Zone}
\tablehead{
\colhead{KOI } &
\colhead{KID} &
\colhead{$T_{\rm eff} (K)$} &
\colhead{$R_* (\rsun)$} &
\colhead{[Fe/H] } &
\colhead{$P$ (Days)} &
\colhead{$R_{\rm P} (\rearth)$} &
\colhead{$F_{\rm P} (F_\oplus)$} 
}
1686.01 & 6149553 & 3414 & 0.30 & -0.1 & 56.87 & 0.95 & 0.30 \\
2418.01 & 10027247 & 3724 & 0.41 & -0.4 & 86.83 & 1.27 & 0.35 \\
\hline
854.01 & 6435936 & 3562 & 0.40 & -0.1 & 56.05 & 1.69 & 0.50 \\
2626.01 & 11768142 & 3482 & 0.35 & -0.1 & 38.10 & 1.37 & 0.66 \\
1422.02 & 11497958 & 3424 & 0.22 & -0.5 & 19.85 & 0.92 & 0.82 \\
\hline
250.04 & 9757613 & 3853 & 0.45 & -0.5 & 46.83 & 1.92 & 1.02 \\
2650.01 & 8890150 & 3735 & 0.40 & -0.5 & 34.99 & 1.18 & 1.15 \\
886.03 & 7455287 & 3579 & 0.33 & -0.4 & 21.00 & 1.14 & 1.47 \\
947.01 & 9710326 & 3717 & 0.43 & -0.3 & 28.60 & 1.84 & 1.61 \\
463.01 & 8845205 & 3504 & 0.34 & -0.2 & 18.48 & 1.80 & 1.70 \\
1879.01 & 8367644 & 3635 & 0.41 & -0.2 & 22.08 & 2.37 & 1.96 
\enddata
\label{tab:hz}
\end{deluxetable*}

\subsection{Planet Occurrence in the Habitable Zone}
\label{ssec:occinhz}
Our final sample contains three planet candidates in the habitable
zone, which is sufficient to allow us to place a lower limit on the
occurrence rate in the habitable zone of late~K and early~M dwarfs. 
We find that planets with the same radii and insolation as 
KOIs~854.01, 1422.02, and 2626.01 could
have been detected around 2853 (73\%), 813 (21\%), and 2131 (55\%) of
the cool dwarfs, respectively. Accordingly, the occurrence rate of
Earth-size ($0.5-1.4\rearth$) planets in the habitable zone is \ehz~planets per star and the occurrence rate of larger ($1.4-4\rearth$)
planets is \sehz~planets per star. We find lower
limits of \hzocclim~Earth-size planets and $\hzoccbig$~$1.4-4\rearth$
planets per cool dwarf habitable zone with 95\% confidence. These 
occurrence rate estimates are most applicable for stars with temperatures 
between $3400$K and $4000$K because 80\% of the stars in our cool dwarf 
sample have temperatures above $3400$K.

\begin{figure}[htbp]
\begin{center}
\centering
\includegraphics[width=0.5\textwidth]{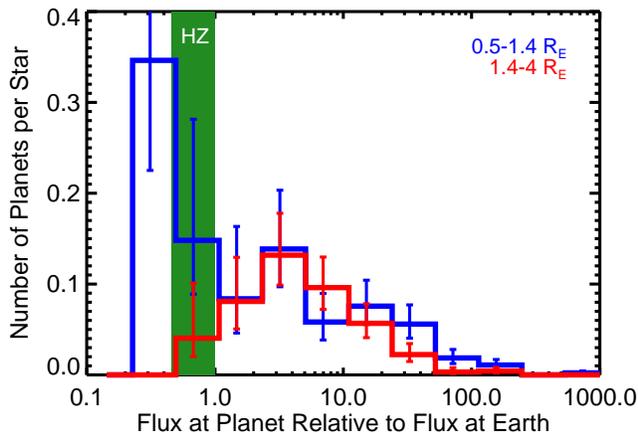}
\end{center}
\caption{Planet occurrence rate versus insolation for Earth-size
planets ($0.5-1.4\rearth$, blue) and $1.4-4\rearth$ planets (red). The
green box marks the habitable zone. The error bars indicate
the errors from binomial statistics and do not include errors from the
stellar and planetary radius estimates although we do consider those errors as discussed in Section~\ref{ssec:occinhz}.}
\label{fig:fluxocc}
\end{figure}

As shown in Figure \ref{fig:fluxocc}, the occurrence rate of
$1.4-4\rearth$ planets peaks for insolation levels $2.2-4.7$~times
higher than that received by the Earth ($F_\oplus$) and falls off at
higher and lower insolation levels. The occurrence rate of Earth-size
planets is roughly constant per logarithmic insolation bin 
for insolation levels between $0.2-50F_\oplus$ and decreases for 
higher levels of insolation. The large error bars at low insolation levels 
should shrink as the \kepler mission continues and becomes 
more sensitive to small planets in longer-period planets.

Our result for the occurrence rate of $1.4-4\rearth$~planets within
the habitable zones of late~K and early~M~dwarfs is lower than the
$42^{+54}_{-13}\%$ occurrence rate reported by
\citet{bonfils_et_al2011b} from an analysis of the HARPS radial
velocity data. The difference between our results may be due in part
to the difficulty of converting measured minimum masses into planetary
radii and the definition of a ``Super Earth'' for both surveys. Small
number statistics may also factor into the difference. \cite{bonfils_et_al2011b} 
surveyed 102~M dwarfs and found two Super Earths within the 
habitable zone: Gl 581c \citep{selsis_et_al2007, vonbloh_et_al2007} 
and Gl~667Cc \citep{anglada-escude_et_al2012, delfosse_et_al2012}. 
Their 42\% estimate of the 
occurrence rate of Super Earths in the habitable zone includes a 
large correction for incompleteness. In comparison, the \kepler sample contains
\ndwarfs M dwarfs with three small habitable zone planets.

\begin{figure}[htbp]
\begin{center}
\centering
\includegraphics[width=0.5\textwidth]{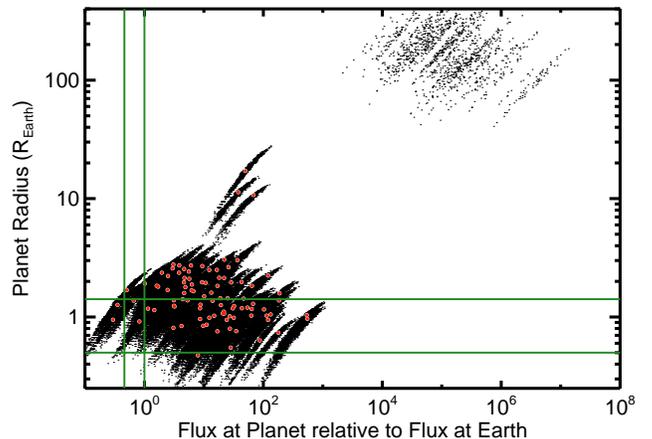}
\end{center}
\caption{Planet radii versus insolation for the population of planet
candidates generated in the perturbation analysis. The best-fit
parameters for each planet candidate are indicated by red circles and
the perturbed realizations are marked by black points. The green lines mark
the boundaries of the habitable zone as defined in Section~\ref{ssec:hz}.}
\label{fig:perturbplanets}
\end{figure}

Due to the small sample size and the need to account for uncertainties in the stellar parameters, we also conduct a perturbation analysis in which we generate 10,000 realizations of each of the \ndwarfs cool dwarfs and recalculate the occurrence rate within the habitable zone for each realization. We generate the population of cool dwarfs by drawing 10,000 model fits for each cool dwarf from the Dartmouth Stellar Models. We weight the probability that a particular model is selected by the likelihoods computed in Section~\ref{sec:methods} so that the population of models for each star represents the probability density function for the stellar parameters. For the planet host stars, we then compute the radii, semimajor axes, and insolation levels of the associated planet candidates. The full population of perturbed planet candidates is plotted in Figure \ref{fig:perturbplanets}. The realization ``ellipses'' are diagonally elongated due to the correlation between stellar temperature and radius.

For each realization of perturbed stars and associated planet
candidates, we calculate the number of cool dwarfs for which each
perturbed planet could have been detected. We report the median
occurrence rates and the 68\% confidence intervals in
Table~\ref{tab:hzfluxocc} as a function of planet radius and
insolation. The estimated occurrence rates resulting from the
perturbation analysis are consistent with the occurrence rates plotted in
Figure~\ref{fig:fluxocc} for the best-fit model parameters.

In addition to refining our estimate of the mean number of planets in
the habitable zone, the perturbation analysis also allows us to
estimate the likelihood that each of the planet candidates lies within
the habitable zone. We find that the most likely habitable planet is
KOI~2626.01, which lies within the habitable zone in 4,907 of the
10,000 realizations. KOIs~2650.01, 1422.02, 250.04, and 947.01 are
also promising candidates and are within the habitable zone in 47\%,
46\%, 28\%, and 22\% of the realizations, respectively. KOIs~886.03,
463.01, 1686.01, 1078.03, 1879.01, 817.01, and 571.04 have much lower
habitability fractions (11\%, 8\%, 7\%, 5\%, 5\%, 3\%, and 2\%) but
still contribute to the overall estimate of the occurrence rate of
planets within the habitable zone of cool dwarfs.

\begin{deluxetable}{lcc}
\tablecaption{Results of Perturbation Analysis: Planet Occurrence Rate as a Function of Flux for Late K and Early M Dwarfs}
\small
\tablecolumns{3}
\tablewidth{0pt}
\tablehead{
\colhead{} &
\multicolumn{2}{c}{Planet Radius} \\
\cline{2-3}
\colhead{Flux $(F_{\rm Earth})$} & 
\colhead{$0.5-1.4 \rearth$} & 
\colhead{$1.4-4 \rearth$}
}
$0.10-0.21$ & --- & --- \\[1.2ex]
$0.21-0.46$ & $0.256^{+0.210}_{-0.142}$ & --- \\[1.2ex]
$0.46-1.00$ & $0.155^{+0.138}_{-0.098}$ & $0.039^{+0.038}_{-0.039}$ \\[1.2ex]
$1.00-2.17$ & $0.153^{+0.089}_{-0.064}$ & $0.084^{+0.033}_{-0.026}$ \\[1.2ex]
$2.17-4.73$ & $0.133^{+0.055}_{-0.043}$ & $0.120^{+0.031}_{-0.030}$ \\[1.2ex]
$4.73-10.27$ & $0.131^{+0.049}_{-0.042}$ & $0.069^{+0.023}_{-0.021}$ \\[1.2ex]
$10.27-22.33$ & $0.100^{+0.025}_{-0.023}$ & $0.043^{+0.013}_{-0.011}$ \\[1.2ex]
$22.33-48.55$ & $0.047^{+0.012}_{-0.012}$ & $0.013^{+0.006}_{-0.008}$ \\[1.2ex]
$48.55-105.55$ & $0.017^{+0.006}_{-0.006}$ & $0.004^{+0.004}_{-0.001}$ \\[1.2ex]
$105.55-229.45$ & $0.007^{+0.003}_{-0.003}$ & $0.002^{+0.001}_{-0.002}$ \\[1.2ex]
$229.45-498.81$ & $0.002^{+0.001}_{-0.002}$ & --- \\[1.2ex]
$498.81-1084.37$ & --- & --- \\[1.2ex]
\enddata
\label{tab:hzfluxocc}
\end{deluxetable}

\section{Summary and Conclusions}
\label{sec:conc}
We update the stellar parameters for the coolest stars in the \kepler
target list by comparing the observed colors of the stars to the
colors of model stars from the Dartmouth Stellar Evolutionary
Program. Our final sample contains \ndwarfs dwarf stars with revised
temperatures cooler than 4000K. In agreement with previous research,
we find that the temperatures and radii of the coolest stars listed in
the KIC are overestimated. For a typical star, our revised estimates
are 130K cooler and 31\% smaller. We also refit the light curves of
the associated planet candidates to better constrain the planet
radius/star radius ratios and combine the revised radius ratios with
the improved stellar radii of the \nhost~host stars to determine the
radii of the \nplanets~planet candidates in our sample.

In the next stage of our analysis, we compute the planet occurrence
rate by comparing the number of planet candidates to the number of
stars around which \kepler could have detected planets with the same
radius and orbital period or insolation. We find that the mean number
of Earth-size ($0.5-1.4\rearth$) planets and $1.4-4\rearth$ planets
with orbital periods shorter than 50~days are \efifty
and \sefifty planets per star, respectively. Our
occurrence rate for $2-4\rearth$ planets is consistent with the value
reported by \citet{howard_et_al2012} and our occurrence rate for 
$2-32\rearth$ planets is slightly lower than the
occurrence rate found by \citet{gaidos_et_al2012}.

The calculated occurrence rate of Earth-size ($0.5-1.4\rearth$)
planets with orbital periods shorter than 50~days is consistent 
with a flat occurrence rate for temperatures below 4000K, but the 
temperature dependence of the occurrence rate of $1.4-4\rearth$
planets is significantly different. We estimate an occurrence rate of 
$0.61^{+0.08}_{-0.06}$ $1.4-4\rearth$ planets per hotter
star (3723K$\leq T_{\rm eff} \leq 4000$K) and 
$0.19^{+0.07}_{-0.05}$ per cooler star (3122K$ \leq T_{\rm eff} <$ 3723K),
noting that 74\% of the stars in the cool group have 
temperatures between $3400$K and $3701$K.
The apparent decline in the $1.4-4\rearth$~planet occurrence rate at cooler
temperatures might be due to the decreased surface density in the
circumstellar disks of very low-mass stars and the longer orbital
timescales at a given separation \citep{laughlin_et_al2004,
adams_et_al2005, ida+lin2005, kennedy+kenyon2008}.

We also estimate the occurrence rate of potentially habitable planets
around cool stars. We find that the occurrence rate of small (0.5
--1.4 $\rearth$) planets within the habitable zone is
\ehz planets per cool dwarf. This result is lower
than the M~dwarf planet occurrence rates found by radial velocity
surveys \citep{bonfils_et_al2011b}, but higher than some estimates of
the occurrence rate for Sunlike stars
\citep[e.g.,][]{cantanzarite+shao2011}. The relatively high occurrence
rate of potentially habitable planets around cool stars bodes well for
future missions to characterize habitable planets because the majority
of the stars in the solar neighborhood are M~dwarfs. Given that there
are \nearlym early M dwarfs within
10~parsecs,\footnote{\url{http://www.chara.gsu.edu/RECONS/census.posted.htm}} 
we estimate that there are at least \nnewplanets~Earth-size planets in
the habitable zones of nearby M~dwarfs that could be discovered by future missions to find nearby  Earth-like planets. Applying a geometric correction for the transit probability and
assuming that the space density of M~dwarfs is uniform, we find that the nearest transiting Earth-size planet in the habitable zone of an
M~dwarf is less than $\distlim$~pc away with 95\% confidence. Removing the requirement that the planet transits, we find that the nearest non-transiting Earth-size planet in the habitable zone is within $\distlimnotransit$~pc with 95\% confidence. The most probable distances
 to the nearest transiting and non-transiting Earth-size planets in the habitable zone are $\distlimmostprob$~pc and $\distlimmostprobnotransit$~pc, respectively.

\begin{acknowledgments}
C.D. is supported by a National Science Foundation Graduate Research Fellowship. We acknowledge support from the \kepler Participating Scientist Program via grant NNX12AC77G. We thank the referee, Philip Muirhead, for providing comments that improved the paper. We acknowledge helpful conversations with S. Ballard, Z. Berta, J. Carter, R. Dawson, J.-M. Desert,  A. Dupree, F. Fressin, A. Howard, J. Irwin, D. Latham, R. Murray-Clay, and G. Torres. We thank R. Kopparapu for correspondence that led to a correction. This paper includes data collected by the \kepler mission. Funding for the Kepler mission is provided by the NASA Science Mission directorate. We thank the \kepler team for acquiring, reducing, and sharing their data. This publication makes use of data products from the Two Micron All Sky Survey, which is a joint project of the University of Massachusetts and the Infrared Processing and Analysis Center/California Institute of Technology, funded by the National Aeronautics and Space Administration and the National Science Foundation. All of the data presented in this paper were obtained from the Mikulski Archive for Space Telescopes (MAST). STScI is operated by the Association of Universities for Research in Astronomy, Inc., under NASA contract NAS5-26555. Support for MAST for non-HST data is provided by the NASA Office of Space Science via grant NNX09AF08G and by other grants and contracts.
\end{acknowledgments}

\newpage

\begin{deluxetable*}{llllrrr}
\small
\tablewidth{0pt}
\tablecolumns{7}
\tablecaption{Revised Cool Star Properties}
\tablehead{
\colhead{KID} & 
\colhead{$T_{\rm eff}$ (K)}& 
\colhead{$R_*(\rsun)$} & 
\colhead{$M_* (\msun)$} & 
\colhead{$\log g$} & 
\colhead{[Fe/H]}  &
\colhead{Dist (pc)}
}
1162635 & 3759$^{+50}_{-50}$ & 0.494$^{+0.05}_{-0.05}$ & 0.505$^{+0.05}_{-0.05}$ & 4.754$^{+0.06}_{-0.06}$ & -0.10$^{+0.1}_{-0.1}$ & 261.3$^{+17}_{-12}$ \\[1.2ex]
1292688 & 3774$^{+77}_{-50}$ & 0.530$^{+0.07}_{-0.05}$ & 0.539$^{+0.06}_{-0.05}$ & 4.722$^{+0.06}_{-0.07}$ & 0.00$^{+0.1}_{-0.1}$ & 282.1$^{+43}_{-10}$ \\[1.2ex]
1293177 & 3385$^{+50}_{-50}$ & 0.216$^{+0.05}_{-0.05}$ & 0.204$^{+0.05}_{-0.05}$ & 5.077$^{+0.06}_{-0.06}$ & -0.40$^{+0.1}_{-0.1}$ & 101.7$^{+15}_{-10}$ \\[1.2ex]
1293393 & 3953$^{+137}_{-54}$ & 0.536$^{+0.13}_{-0.05}$ & 0.555$^{+0.12}_{-0.05}$ & 4.725$^{+0.06}_{-0.13}$ & -0.20$^{+0.4}_{-0.1}$ & 454.2$^{+130}_{-31}$ \\[1.2ex]
1429729 & 3903$^{+76}_{-60}$ & 0.523$^{+0.07}_{-0.05}$ & 0.541$^{+0.07}_{-0.05}$ & 4.735$^{+0.06}_{-0.07}$ & -0.20$^{+0.2}_{-0.1}$ & 380.0$^{+61}_{-32}$ \\[1.2ex]
1430893 & 3929$^{+98}_{-58}$ & 0.541$^{+0.07}_{-0.05}$ & 0.564$^{+0.07}_{-0.05}$ & 4.724$^{+0.06}_{-0.06}$ & -0.10$^{+0.2}_{-0.1}$ & 269.8$^{+42}_{-20}$ \\[1.2ex]
1433760 & 3296$^{+50}_{-50}$ & 0.213$^{+0.05}_{-0.05}$ & 0.196$^{+0.05}_{-0.05}$ & 5.072$^{+0.06}_{-0.06}$ & -0.10$^{+0.1}_{-0.1}$ & 109.8$^{+13}_{-13}$ \\[1.2ex]
1569682 & 3860$^{+93}_{-78}$ & 0.514$^{+0.06}_{-0.07}$ & 0.544$^{+0.07}_{-0.05}$ & 4.752$^{+0.07}_{-0.06}$ & -0.10$^{+0.2}_{-0.1}$ & 262.8$^{+41}_{-44}$ \\[1.2ex]
1569863 & 3591$^{+50}_{-53}$ & 0.360$^{+0.05}_{-0.06}$ & 0.384$^{+0.07}_{-0.06}$ & 4.910$^{+0.07}_{-0.06}$ & -0.30$^{+0.1}_{-0.1}$ & 157.0$^{+25}_{-29}$ \\[1.2ex]
1572802 & 3878$^{+53}_{-88}$ & 0.535$^{+0.05}_{-0.06}$ & 0.545$^{+0.05}_{-0.06}$ & 4.719$^{+0.06}_{-0.06}$ & -0.10$^{+0.1}_{-0.1}$ & 246.4$^{+25}_{-36}$ \\[1.2ex]
\enddata
\label{tab:stubcoolstars}
\tablecomments{Table \ref{tab:stubcoolstars} is published in its entirety in the electronic edition. A portion is shown here for guidance regarding its form and content.}
\end{deluxetable*}

\LongTables
\begin{deluxetable*}{lllllllllll}
\footnotesize
\tablecolumns{11}
\tablewidth{0.5\linewidth}
\tablecaption{Revised Properties for Planet Candidates Orbiting Small Stars}
\tablehead{
\colhead{KOI}& 
\colhead{KID} & 
\colhead{t0 (Days)} & 
\colhead{$P$ (Days)} & 
\colhead{$a/R_*$\tablenotemark{a}} & 
\colhead{$R_p/R_*$} & 
\colhead{$b$} & 
\colhead{$R_{\rm P} (\rearth)$} & 
\colhead{$F (F_\oplus)$} &
\colhead{$T_{\rm eff} (K) $} & 
\colhead{$R_* (\rsun)$}
}
\startdata
 247.01 &  11852982 &    1.525 &   13.815 &  47.536 &  0.030 &    0.4 & $1.41^{+0.26}_{-0.29}$ & $4.41_{-2.95}^{+5.61}$ &    3725 &  0.437 \\
 248.01\tablenotemark{b} &   5364071 &    4.593 &    7.028 &  17.897 &  0.032 &    0.6 & $1.83^{+0.18}_{-0.26}$ & $16.90_{-1.88}^{+3.57}$ &    3903 &  0.523 \\
 248.02\tablenotemark{c} &   5364071 &    6.158 &   10.913 &  21.948 &  0.047 &    0.8 & $2.69^{+0.26}_{-0.38}$ & $9.40_{-6.67}^{+9.86}$ &    3903 &  0.523 \\
 248.03 &   5364071 &    2.076 &    2.577 &  10.121 &  0.032 &    0.5 & $1.83^{+0.18}_{-0.26}$ & $64.39_{-3.94}^{+5.83}$ &    3903 &  0.523 \\
 248.04 &   5364071 &   11.080 &   18.596 &  51.184 &  0.034 &    0.5 & $1.96^{+0.19}_{-0.27}$ & $4.62_{-2.14}^{+3.62}$ &    3903 &  0.523 \\
 249.01 &   9390653 &    3.871 &    9.549 &  44.353 &  0.040 &    0.3 & $1.60^{+0.22}_{-0.22}$ & $4.65_{-6.22}^{+7.56}$ &    3514 &  0.370 \\
 250.01\tablenotemark{d} &   9757613 &   10.720 &   12.283 &  34.265 &  0.056 &    0.3 & $2.73^{+0.63}_{-0.54}$ & $6.06_{-3.46}^{+4.21}$ &    3853 &  0.447 \\
 250.02\tablenotemark{e} &   9757613 &   11.877 &   17.251 &  62.567 &  0.056 &    0.5 & $2.73^{+0.63}_{-0.54}$ & $3.85_{-23.70}^{+28.82}$ &    3853 &  0.447 \\
 250.03 &   9757613 &    1.594 &    3.544 &  11.511 &  0.020 &    0.5 & $0.98^{+0.23}_{-0.19}$ & $31.79_{-1.70}^{+2.07}$ &    3853 &  0.447 \\
 250.04 &   9757613 &   43.087 &   46.828 & 157.259 &  0.039 &    0.6 & $1.92^{+0.44}_{-0.38}$ & $1.02_{-1.68}^{+2.18}$ &    3853 &  0.447 \\
 251.01 &  10489206 &    0.347 &    4.164 &  12.214 &  0.049 &    0.7 & $2.63^{+0.27}_{-0.34}$ & $26.00_{-15.49}^{+29.45}$ &    3743 &  0.488 \\
 251.02 &  10489206 &    0.157 &    5.775 &  18.612 &  0.014 &    0.5 & $0.76^{+0.08}_{-0.10}$ & $16.81_{-0.50}^{+0.94}$ &    3743 &  0.488 \\
 252.01 &  11187837 &   12.059 &   17.605 &  33.315 &  0.045 &    0.5 & $2.37^{+0.25}_{-0.30}$ & $3.82_{-8.78}^{+9.79}$ &    3770 &  0.479 \\
 253.01 &  11752906 &    4.643 &    6.383 &  17.910 &  0.049 &    0.8 & $3.05^{+0.27}_{-0.47}$ & $21.99_{-5.68}^{+6.33}$ &    3919 &  0.574 \\
 254.01\tablenotemark{f} &   5794240 &    1.410 &    2.455 &  11.223 &  0.179 &    0.5 & $10.74^{+0.98}_{-1.44}$ & $68.37_{-1.30}^{+1.67}$ &    3837 &  0.550 \\
 255.01 &   7021681 &   24.694 &   27.522 &  51.142 &  0.045 &    0.3 & $2.77^{+0.24}_{-0.34}$ & $3.07_{-8.91}^{+8.92}$ &    3907 &  0.570 \\
 256.01\tablenotemark{v} &  11548140 &    0.200 &    1.379 &   4.825 &  0.454 &    1.2 & $17.12^{+2.48}_{-2.48}$ & $49.78_{-24.90}^{+24.30}$ &    3410 &  0.346 \\
 463.01 &   8845205 &    0.491 &   18.478 &  69.231 &  0.049 &    0.5 & $1.80^{+0.33}_{-0.38}$ & $1.70_{-1.05}^{+1.02}$ &    3504 &  0.340 \\
 531.01 &  10395543 &    1.255 &    3.687 &  14.775 &  0.089 &    0.9 & $4.77^{+0.63}_{-0.69}$ & $36.55_{-18.38}^{+27.40}$ &    3898 &  0.490 \\
 571.01 &   8120608 &    7.166 &    7.267 &  22.444 &  0.025 &    0.4 & $1.37^{+0.14}_{-0.21}$ & $13.98_{-0.83}^{+1.36}$ &    3820 &  0.500 \\
 571.02 &   8120608 &    3.440 &   13.343 &  25.894 &  0.031 &    0.7 & $1.68^{+0.17}_{-0.25}$ & $6.22_{-13.83}^{+19.62}$ &    3820 &  0.500 \\
 571.03 &   8120608 &    1.184 &    3.887 &  12.801 &  0.023 &    0.6 & $1.24^{+0.12}_{-0.19}$ & $32.20_{-5.33}^{+6.16}$ &    3820 &  0.500 \\
 571.04 &   8120608 &   19.360 &   22.407 &  46.240 &  0.025 &    0.5 & $1.39^{+0.14}_{-0.21}$ & $3.12_{-2.37}^{+2.74}$ &    3820 &  0.500 \\
 596.01 &  10388286 &    0.496 &    1.683 &   8.565 &  0.025 &    0.4 & $1.17^{+0.14}_{-0.16}$ & $63.67_{-12.27}^{+14.19}$ &    3626 &  0.430 \\
 739.01 &  10386984 &    1.214 &    1.287 &   6.023 &  0.026 &    0.5 & $1.58^{+0.19}_{-0.14}$ & $187.26_{-1.19}^{+1.37}$ &    3994 &  0.554 \\
 781.01 &  11923270 &    6.418 &   11.598 &  29.230 &  0.055 &    0.7 & $2.54^{+0.30}_{-0.30}$ & $4.65_{-22.44}^{+28.30}$ &    3603 &  0.423 \\
 817.01 &   4725681 &   18.439 &   23.968 &  42.743 &  0.033 &    0.5 & $1.69^{+0.20}_{-0.20}$ & $2.45_{-50.98}^{+87.00}$ &    3758 &  0.474 \\
 817.02 &   4725681 &    3.063 &    8.296 &  45.449 &  0.029 &    0.5 & $1.49^{+0.18}_{-0.17}$ & $10.09_{-1.47}^{+1.99}$ &    3758 &  0.474 \\
 818.01 &   4913852 &    5.940 &    8.114 &  25.959 &  0.038 &    0.4 & $1.65^{+0.21}_{-0.21}$ & $6.67_{-0.78}^{+1.20}$ &    3564 &  0.401 \\
 854.01 &   6435936 &   33.001 &   56.055 &  90.045 &  0.039 &    0.4 & $1.69^{+0.33}_{-0.21}$ & $0.50_{-3.22}^{+4.96}$ &    3562 &  0.400 \\
 886.01\tablenotemark{g} &   7455287 &    1.978 &    8.011 &   6.286 &  0.038 &    1.1 & $1.38^{+0.30}_{-0.27}$ & $5.30_{-2.19}^{+3.12}$ &    3579 &  0.330 \\
 886.02\tablenotemark{h} &   7455287 &   10.709 &   12.072 &   6.370 &  0.023 &    1.3 & $0.81^{+0.18}_{-0.16}$ & $3.07_{-0.18}^{+0.35}$ &    3579 &  0.330 \\
 886.03\tablenotemark{v} &   7455287 &    5.355 &   20.995 &  39.246 &  0.032 &    0.8 & $1.14^{+0.25}_{-0.22}$ & $1.47_{-2.48}^{+4.93}$ &    3579 &  0.330 \\
 898.01\tablenotemark{i} &   7870390 &    9.615 &    9.770 &  27.672 &  0.042 &    0.4 & $2.49^{+0.23}_{-0.23}$ & $12.33_{-1.44}^{+2.85}$ &    3989 &  0.544 \\
 898.02 &   7870390 &    2.032 &    5.170 &  16.115 &  0.033 &    0.5 & $1.96^{+0.18}_{-0.18}$ & $28.81_{-0.69}^{+1.36}$ &    3989 &  0.544 \\
 898.03\tablenotemark{j} &   7870390 &    7.354 &   20.090 &  41.819 &  0.036 &    0.4 & $2.14^{+0.20}_{-0.20}$ & $4.71_{-3.18}^{+4.32}$ &    3989 &  0.544 \\
 899.01 &   7907423 &    3.596 &    7.114 &  23.515 &  0.028 &    0.5 & $1.27^{+0.15}_{-0.25}$ & $8.74_{-7.43}^{+10.10}$ &    3587 &  0.410 \\
 899.02 &   7907423 &    2.114 &    3.307 &  12.885 &  0.021 &    0.4 & $0.95^{+0.12}_{-0.19}$ & $24.26_{-1.22}^{+1.65}$ &    3587 &  0.410 \\
 899.03 &   7907423 &    9.085 &   15.368 &  31.920 &  0.028 &    0.8 & $1.24^{+0.15}_{-0.24}$ & $3.13_{-4.23}^{+4.74}$ &    3587 &  0.410 \\
 936.01 &   9388479 &    7.990 &    9.468 &  27.967 &  0.044 &    0.4 & $1.79^{+0.24}_{-0.26}$ & $4.88_{-11.74}^{+13.16}$ &    3518 &  0.370 \\
 936.02 &   9388479 &    0.580 &    0.893 &   5.775 &  0.025 &    0.4 & $1.03^{+0.14}_{-0.15}$ & $113.74_{-1.51}^{+1.70}$ &    3518 &  0.370 \\
 947.01 &   9710326 &   18.333 &   28.599 &  46.796 &  0.039 &    0.7 & $1.84^{+0.35}_{-0.26}$ & $1.61_{-1.93}^{+2.31}$ &    3717 &  0.430 \\
 952.01\tablenotemark{k} &   9787239 &    0.274 &    5.901 &  19.376 &  0.039 &    0.4 & $2.15^{+0.28}_{-0.28}$ & $18.06_{-44.92}^{+53.82}$ &    3787 &  0.506 \\
 952.02\tablenotemark{l} &   9787239 &    4.351 &    8.752 &  19.985 &  0.035 &    0.7 & $1.94^{+0.25}_{-0.26}$ & $10.68_{-0.61}^{+1.16}$ &    3787 &  0.506 \\
 952.03 &   9787239 &   18.525 &   22.780 &  48.891 &  0.047 &    0.4 & $2.58^{+0.33}_{-0.34}$ & $2.98_{-1.10}^{+1.63}$ &    3787 &  0.506 \\
 952.04 &   9787239 &    0.400 &    2.896 &  13.641 &  0.026 &    0.5 & $1.43^{+0.18}_{-0.19}$ & $46.65_{-17.22}^{+25.47}$ &    3787 &  0.506 \\
1078.01 &  10166274 &    0.720 &    3.354 &  16.129 &  0.035 &    0.4 & $1.97^{+0.24}_{-0.25}$ & $43.04_{-14.85}^{+21.56}$ &    3878 &  0.523 \\
1078.02 &  10166274 &    1.417 &    6.877 &  20.830 &  0.044 &    0.9 & $2.50^{+0.30}_{-0.31}$ & $16.52_{-5.70}^{+8.28}$ &    3878 &  0.523 \\
1078.03 &  10166274 &   15.729 &   28.463 &  71.424 &  0.039 &    0.5 & $2.22^{+0.27}_{-0.28}$ & $2.49_{-0.86}^{+1.25}$ &    3878 &  0.523 \\
1085.01 &  10118816 &    0.219 &    7.718 &  26.930 &  0.018 &    0.6 & $1.02^{+0.11}_{-0.10}$ & $14.89_{-4.05}^{+6.29}$ &    3878 &  0.535 \\
1141.01 &   8346392 &    3.424 &    5.728 &  17.940 &  0.024 &    0.5 & $1.44^{+0.16}_{-0.14}$ & $24.99_{-7.43}^{+10.78}$ &    3976 &  0.550 \\
1146.01 &   8351704 &    1.504 &    7.097 &  23.314 &  0.019 &    0.4 & $0.99^{+0.13}_{-0.10}$ & $12.32_{-3.67}^{+5.98}$ &    3778 &  0.470 \\
1164.01 &  10341831 &    0.780 &    0.934 &   1.768 &  0.014 &    0.3 & $0.74^{+0.08}_{-0.08}$ & $178.11_{-52.00}^{+68.87}$ &    3711 &  0.475 \\
1201.01 &   4061149 &    0.690 &    2.758 &  18.588 &  0.023 &    0.4 & $1.19^{+0.18}_{-0.16}$ & $43.54_{-16.58}^{+27.86}$ &    3728 &  0.482 \\
1393.01 &   9202151 &    1.164 &    1.695 &   7.709 &  0.037 &    0.4 & $2.24^{+0.20}_{-0.30}$ & $120.22_{-43.59}^{+41.73}$ &    3872 &  0.563 \\
1397.01 &   9427402 &    0.829 &    6.247 &  30.153 &  0.036 &    0.4 & $2.14^{+0.20}_{-0.20}$ & $21.87_{-5.95}^{+8.08}$ &    3957 &  0.542 \\
1422.01 &  11497958 &    1.568 &    5.842 &  22.474 &  0.035 &    0.4 & $0.84^{+0.19}_{-0.19}$ & $4.20_{-2.16}^{+3.80}$ &    3424 &  0.220 \\
1422.02 &  11497958 &   14.559 &   19.850 &  51.985 &  0.038 &    0.4 & $0.92^{+0.21}_{-0.21}$ & $0.82_{-0.42}^{+0.74}$ &    3424 &  0.220 \\
1422.03 &  11497958 &    0.743 &    3.622 &   7.933 &  0.020 &    0.9 & $0.47^{+0.11}_{-0.11}$ & $7.95_{-4.08}^{+7.18}$ &    3424 &  0.220 \\
1427.01 &  11129738 &    2.463 &    2.613 &   9.757 &  0.023 &    0.5 & $1.29^{+0.12}_{-0.16}$ & $67.13_{-21.75}^{+25.15}$ &    3979 &  0.523 \\
1649.01 &  11337141 &    2.239 &    4.044 &   7.983 &  0.019 &    0.9 & $1.02^{+0.11}_{-0.15}$ & $27.15_{-10.05}^{+11.74}$ &    3767 &  0.479 \\
1681.01 &   5531953 &    6.486 &    6.939 &  15.493 &  0.027 &    0.8 & $1.18^{+0.18}_{-0.15}$ & $8.63_{-2.94}^{+4.77}$ &    3608 &  0.400 \\
1686.01 &   6149553 &   43.529 &   56.867 & 102.482 &  0.029 &    0.5 & $0.95^{+0.16}_{-0.16}$ & $0.30_{-0.12}^{+0.19}$ &    3414 &  0.300 \\
1702.01 &   7304449 &    1.082 &    1.538 &   9.008 &  0.028 &    0.6 & $0.80^{+0.15}_{-0.15}$ & $27.41_{-12.57}^{+20.88}$ &    3304 &  0.260 \\
1843.01 &   5080636 &    4.103 &    4.195 &  19.152 &  0.026 &    0.4 & $1.26^{+0.14}_{-0.22}$ & $19.30_{-8.02}^{+9.46}$ &    3584 &  0.450 \\
1843.02 &   5080636 &    4.025 &    6.356 &  38.543 &  0.018 &    0.5 & $0.86^{+0.10}_{-0.15}$ & $11.09_{-4.61}^{+5.43}$ &    3584 &  0.450 \\
1867.01 &   8167996 &    0.033 &    2.550 &   9.819 &  0.022 &    0.5 & $1.20^{+0.12}_{-0.13}$ & $53.87_{-16.96}^{+24.11}$ &    3799 &  0.492 \\
1867.02 &   8167996 &    6.446 &   13.969 &  26.759 &  0.045 &    1.0 & $2.42^{+0.25}_{-0.27}$ & $5.58_{-1.76}^{+2.50}$ &    3799 &  0.492 \\
1867.03 &   8167996 &    2.404 &    5.212 &  15.672 &  0.020 &    0.5 & $1.07^{+0.11}_{-0.12}$ & $20.76_{-6.54}^{+9.29}$ &    3799 &  0.492 \\
1868.01 &   6773862 &   13.183 &   17.761 &  76.082 &  0.034 &    0.4 & $2.10^{+0.19}_{-0.20}$ & $5.68_{-1.65}^{+2.16}$ &    3950 &  0.560 \\
1879.01 &   8367644 &    2.731 &   22.085 &  69.891 &  0.053 &    0.5 & $2.37^{+0.38}_{-0.35}$ & $1.96_{-0.75}^{+1.21}$ &    3635 &  0.410 \\
1880.01 &  10332883 &    0.847 &    1.151 &   5.801 &  0.024 &    0.7 & $1.38^{+0.13}_{-0.22}$ & $182.97_{-72.47}^{+69.88}$ &    3855 &  0.530 \\
1907.01 &   7094486 &    9.197 &   11.350 &  32.483 &  0.033 &    0.5 & $1.96^{+0.19}_{-0.18}$ & $9.30_{-2.52}^{+3.90}$ &    3901 &  0.542 \\
2006.01 &  10525027 &    0.233 &    3.273 &  12.574 &  0.015 &    0.5 & $0.76^{+0.14}_{-0.11}$ & $35.00_{-13.95}^{+23.72}$ &    3809 &  0.455 \\
2036.01 &   6382217 &    7.635 &    8.411 &  27.409 &  0.028 &    0.4 & $1.60^{+0.15}_{-0.30}$ & $13.30_{-5.94}^{+6.10}$ &    3903 &  0.523 \\
2036.02 &   6382217 &    3.489 &    5.795 &  19.205 &  0.019 &    0.6 & $1.07^{+0.10}_{-0.20}$ & $21.85_{-9.76}^{+10.02}$ &    3903 &  0.523 \\
2057.01 &   9573685 &    3.200 &    5.945 &  18.668 &  0.019 &    0.5 & $1.11^{+0.10}_{-0.14}$ & $21.67_{-7.63}^{+7.87}$ &    3900 &  0.537 \\
2058.01 &  10329835 &    0.575 &    1.524 &   8.046 &  0.018 &    0.5 & $1.05^{+0.10}_{-0.13}$ & $133.13_{-46.43}^{+47.95}$ &    3900 &  0.537 \\
2090.01 &  11348997 &    3.845 &    5.132 &  23.462 &  0.027 &    0.4 & $1.44^{+0.15}_{-0.22}$ & $18.90_{-7.35}^{+8.02}$ &    3688 &  0.497 \\
2130.01 &   2161536 &    3.445 &   16.855 &  50.586 &  0.031 &    0.4 & $1.88^{+0.17}_{-0.25}$ & $6.27_{-2.17}^{+2.21}$ &    3972 &  0.565 \\
2156.01 &   2556650 &    0.835 &    2.852 &   9.813 &  0.223 &    1.2 & $11.32^{+1.22}_{-1.29}$ & $37.83_{-11.55}^{+14.90}$ &    3694 &  0.464 \\
2179.01 &  10670119 &    3.851 &   14.871 &  43.731 &  0.027 &    0.4 & $1.23^{+0.15}_{-0.18}$ & $3.20_{-1.21}^{+1.44}$ &    3591 &  0.410 \\
2179.02 &  10670119 &    1.564 &    2.733 &  21.112 &  0.026 &    0.5 & $1.18^{+0.14}_{-0.17}$ & $30.65_{-11.59}^{+13.80}$ &    3591 &  0.410 \\
2191.01 &   5601258 &    7.441 &    8.848 &  29.479 &  0.019 &    0.6 & $0.96^{+0.11}_{-0.13}$ & $8.61_{-3.00}^{+4.10}$ &    3724 &  0.460 \\
2238.01 &   8229458 &    0.313 &    1.647 &   8.090 &  0.016 &    0.5 & $0.95^{+0.09}_{-0.09}$ & $120.04_{-34.30}^{+38.46}$ &    3900 &  0.537 \\
2306.01 &   6666233 &    0.040 &    0.512 &   3.419 &  0.018 &    0.5 & $1.04^{+0.10}_{-0.15}$ & $538.37_{-193.75}^{+219.33}$ &    3878 &  0.520 \\
2329.01 &  11192235 &    0.326 &    1.615 &   9.104 &  0.021 &    0.6 & $1.16^{+0.12}_{-0.12}$ & $102.58_{-29.72}^{+41.70}$ &    3815 &  0.498 \\
2347.01 &   8235924 &    0.352 &    0.588 &   3.717 &  0.016 &    0.4 & $0.97^{+0.09}_{-0.09}$ & $550.32_{-145.42}^{+187.35}$ &    3972 &  0.565 \\
2418.01 &  10027247 &   15.600 &   86.830 & 116.837 &  0.028 &    0.5 & $1.27^{+0.24}_{-0.17}$ & $0.35_{-0.13}^{+0.24}$ &    3724 &  0.414 \\
2453.01 &   8631751 &    0.235 &    1.531 &  14.100 &  0.024 &    0.5 & $1.03^{+0.23}_{-0.18}$ & $62.60_{-28.77}^{+57.58}$ &    3565 &  0.400 \\
2542.01 &   6183511 &    0.000 &    0.727 &   4.643 &  0.020 &    0.4 & $0.63^{+0.11}_{-0.17}$ & $87.24_{-48.09}^{+74.20}$ &    3339 &  0.288 \\
2626.01 &  11768142 &   25.703 &   38.098 &  36.283 &  0.036 &    0.9 & $1.37^{+0.43}_{-0.21}$ & $0.66_{-0.30}^{+0.78}$ &    3482 &  0.350 \\
2650.01 &   8890150 &    4.280 &   34.987 &  54.052 &  0.027 &    0.5 & $1.18^{+0.40}_{-0.15}$ & $1.15_{-0.47}^{+1.53}$ &    3735 &  0.400 \\
2650.02 &   8890150 &    2.155 &    7.054 &  30.813 &  0.019 &    0.5 & $0.84^{+0.29}_{-0.11}$ & $9.73_{-3.98}^{+12.94}$ &    3735 &  0.400 \\
2662.01 &   3426367 &    0.742 &    2.104 &  13.578 &  0.015 &    0.5 & $0.55^{+0.08}_{-0.08}$ & $28.22_{-10.41}^{+16.17}$ &    3410 &  0.345 \\
\enddata
\tablenotetext{a}{This column lists the ratio estimated from the fit to the light curve. We compute the geometric probability of transit using the semimajor axis determined from the planet orbital period and the host star mass listed in Table~\ref{tab:stubcoolstars}. }
\tablenotetext{b}{Kepler-49b \citep{steffen_et_al2012b, xie2012}}
 \tablenotetext{c}{Kepler-49c \citep{steffen_et_al2012b, xie2012}}
 \tablenotetext{d}{Kepler-26b \citep{steffen_et_al2012}}
\tablenotetext{e}{Kepler-26c \citep{steffen_et_al2012}}
 \tablenotetext{f}{Confirmed by \citet{johnson_et_al2012}}
 \tablenotetext{g}{Kepler-54b \citep{steffen_et_al2012b}}
  \tablenotetext{h}{Kepler-54c \citep{steffen_et_al2012b}}
 \tablenotetext{i}{Confirmed by \citet{xie2012}}
 \tablenotetext{j}{Confirmed by \citet{xie2012}}
 \tablenotetext{k}{Kepler-32b \citep{fabrycky_et_al2012b}}
 \tablenotetext{l}{Kepler-32c \citep{fabrycky_et_al2012b}}
 \tablenotetext{v}{Transits noted as ``v''-shaped by \citet{batalha_et_al2012}}
 \label{tab:coolkois}
\end{deluxetable*}

\end{document}